\documentclass[twocolumn, showpacs, preprintnumbers,superscriptaddress, amsmath, amssymb, pre]{revtex4-2}
\usepackage{graphicx}
\usepackage{dcolumn}
\usepackage{bm}
\usepackage{epstopdf}
\usepackage{cleveref}

\usepackage{color}
\usepackage[normalem]{ulem}

\newcommand{\Pm}[2]{P_{-,(#1)(#2)}}
\newcommand{\Qm}[2]{Q_{-,(#1)(#2)}}

\begin{document}

\title{
Green's Function-Free Formalism of Projective Truncation Approximation}

\author{Kou-Han Ma}
\affiliation{International Center for Quantum Materials and School of Physics, Peking University, Beijing 100871, China}

\author{Yue-Hong Wu}
\affiliation{School of Physics and Key Laboratory of Quantum State Construction and Manipulation (Ministry of Education), Renmin University of China, Beijing 100872, China}

\author{Ning-Hua Tong}
\email{nhtong@ruc.edu.cn}
\affiliation{School of Physics and Key Laboratory of Quantum State Construction and Manipulation (Ministry of Education), Renmin University of China, Beijing 100872, China}

\date{\today}

\begin{abstract}
In previous works, the projected truncation approximation (PTA) was developed as a systematic and controlled method to truncate the equation of motion of Green's functions (GFs) for a given quantum or classical many-body Hamiltonian. The static averages are obtained self-consistently with the GF through the spectral theorem. In this work, PTA is reformulated as a self-consistent theory for the reduced density matrices (RDMs) without reference to GF. We separately discuss the issues of determining the dynamical matrix ${\bf M}$ and solving the physical quantities from it. The properties of ${\bf M}$ is clarified and the solution of PTA equations is cast into an over-constrained optimization problem. This makes connection of the present theory to the variational RDM theory. We discuss various issues of PTA under this formalism, including the scheme of alternative inner product, the generalized virial theorem, the generalized Wick's theorem, and the static component problem of PTA. 
\end{abstract}


\maketitle

\begin{section}{Introduction}

The many-body correlation is the source of many basic or fascinating phenomena in the field of condensed matter physics, quantum chemistry, nuclear physics, {\it etc}. Green's function (GF) is one of the frequently used theoretical tools for studying the correlation effect\cite{fetter2012quantum,aryasetiawan1998gw,onida2002electronic}. Among the many different formalisms of the equilibrium state GF, theories based on the equation of motion and the idea of operator projection have been developed since 1960s \cite{mori1965transport,mori1965continued, tserkovnikov1981method,lee1982orthogonal,florencio1987,viswanath1994recursion,kakehashi2004coherent,kakehashi2004self,mancini2004hubbard,fan2018projective,ma2022projective}. Similar idea has been used in the study of ground state as well as excitations in nuclear physics models \cite{rowe1968eom,rowe1968methods,ring1980nuclear} and quantum chemistry \cite{catara1996towards,chatterjee2012excitation}, leading to various levels of random phase approximation (RPA) and Bethe-Salpeter equation (BSE). In recent years, a systematic implementation of the operator projection,  projective truncation approximation (PTA), is carried out to the equation of motion (EOM) of two-time GF. PTA is applied to both quantum and classical many-body systems, with an extendable framework and controllable precision \cite{fan2018projective,fan2019controllable,ma2021interacting,ma2022projective,fan2022spatial,jia2025thermal,wu2025self}.

The PTA is implemented in the following steps. One first chooses a set of basis operators that is relevant to the problem of interest. Then one explicitly writes down the Heisenberg EOM of these operators. For an incomplete set of operators, the EOM is not closed since new operators are generated by the commutator of the basis operator and the Hamiltonian of the system. Choosing a well-defined operator inner product, one can project the newly generated operators into the subspace spanned by the basis operators and effectively close the EOM. Employing this truncation to the EOM of the GF matrix of the basis operators, one can obtain the formal expression of the GF matrix, which, via the spectral theorem, produces a set of equations for the averages involved in the GF EOM and the projection process. Finally, these equations for the averages are solved iteratively, giving simultaneously the solutions to static averages and the GFs. This scheme is based on the GF EOM and will be called the GF-based PTA formalism in the following. 

The above-stated PTA has gained much progress in recent years. For examples, it has been used to study the Kondo problem \cite{fan2018projective} and it reveals the nature of the Kondo screening cloud around the magnetic impurity coupled to a ferromagnetic electrode \cite{fan2022spatial}. It gives out a comprehensive phase diagram of the two-dimensional spinless fermion model \cite{ma2021interacting} that goes beyond the result of traditional random phase approximation (RPA). PTA has been generalized to classical statistical models and used to explore the dispersion \cite{ma2022projective} and thermal damping of phonons \cite{jia2025thermal} in the one-dimensional nonlinear lattice. For classical systems, PTA coincides in spirit to the Koopman operator theory widely used in the nonlinear dynamics study and engineering \cite{brunton2022modern,mauroy2020Koopman}.  Recently, a systematic theory of self-consistent RPA is derived within the framework of PTA, under a particular choice of the basis operators (the particle-hole excitation operators) and the inner product (commutator inner product) \cite{wu2025self}. The Rowe formalism of RPA \cite{rowe1968eom,schuck2021equation} is recovered as the special case of zero temperature. The flexibility of choosing basis operators and the inner product in PTA offers the potential to improve the present RPA.

With the above progress, the GF-based PTA formalism still has some issues to be further explored. First, in this formalism, the inner product must be chosen such that it is consistent with the type of the GF used, in order to guarantee the symmetry of the GF matrix. As a result, the flexibility of choosing the inner product in PTA has never been realized. This is especially a serious issue for RPA since the commutator inner product used to derive RPA allows for zero length operator, which may spoil the applicability of the theory in the high symmetry phase. Second, sometimes the equations for the averages is difficult to solve self-consistently. For example, in the self-consistent RPA \cite{wu2025self}, the iterative process for the solution does not converge properly in the strong coupling regime and is often accompanied with the violation of positive-/negative- definiteness of certain matrices. There are other issues as well, such as the static operator component issue, physical constraints to stabilize the solution, and higher-order averages, etc. In this work, we aim to address these issues by developing a GF-free formalism of PTA.

Based on the GF EOM, the existing GF-based PTA can be more properly regarded as a dynamic approach that naturally targets the excitations and dynamical properties of the studied system such as the time correlation function and spectral function. In contrast, the state-targeting methods, such as the numerically exact diagonalization\cite{dagotto1994correlated,weisse2008exact}, density matrix renormalization group\cite{white1992density,schollwock2011density}, tensor network renormalization group\cite{levin2007tensor,evenbly2015tensor}, and the variational wave function methods, aim to find the best approximation for the target state. The static quantities such as the ground state energy and the static correlation functions are produced directly from these methods, but the excitation and dynamical properties are obtained indirectly. Among them, the variational reduced density matrix (RDM) method gains much progress recently\cite{mazziotti2004realization,mazziotti2007,eugene2024variational,PhysRevA.103.052202,mazziotti2023quantum,PhysRevB.110.035110,kwxs-xnhv}. In the variational RDM method, the ground state energy of the system is rigorously expressed in terms of the one- and two-body RDMs. By varying the ground state energy with respect to RDMs under the physical constraints (usually incomplete), the approximate energy and RDMs of the ground state can be obtained.

In this work, we will reformulate the previous GF-based PTA into a theory of RDMs without directly invoking the GF and its EOM. This reformulation effectively transforms PTA from a type of dynamic theory into a state-targeting method. Thus, it gains the potential to combine the merits of both types of theory. In this GF-free formalism, we start from a linear truncation ansatz for the Heisenberg EOM of a preselected set of operators. A set of constraints for certain elements of RDMs is then derived from the truncated EOM, bypassing the use of spectral theorem. Combined with physical constraints of RDMs, such as the algebraic relations and the N-representability constraints, one gets a set of self-consistent equations for certain elements of RDMs, from which the approximate solution of the RDMs can be obtained. The dynamical quantities, such as GF, spectral function, and the time correlation functions, can be produced from the results of the RDMs, with all the physical requirements automatically fulfilled, such as the causality, diagonal positivity, and the conservation laws. 

The purpose of such a formalism of PTA is towards a deeper understanding of the mathematical structure of the PTA, by which more flexible approximations (such as choosing different inner product) and more stable calculation are attainable. It also bridges the static and dynamical quantities coherently, thereby allows for calculating excitations and dynamical quantities from the results of the restate-based methods such as the variational RDM theory. 

The plan of this paper is as follows.
In Sec.\ref{Sec2}, we establish the general formalism of the GF-free PTA. We start from the ansatz of linearly truncated operator EOM where the dynamic matrix \({\bf M}\) is introduced. We derive the relations and constraints that certain basic averages need to fulfill. Expressions for the GFs and spectral functions of the basis operators are given in terms of the RDMs.
In Sec.\ref{Sec3}, we analyze the physical constraints that the thermodynamic quantities must satisfy, including the N-representability conditions of RDMs, conservation laws, and restrictions imposed by operator algebra, among others. 
In Sec.\ref{Sec4}, we address issue of determining the dynamic matrix \({\bf M}\), with particular emphasis on the operator projection, thereby completing the formulation of the GF-free PTA as a fully self-consistent theory.
In Sec.\ref{Sec5}, we derive reduction formulas for higher-order quantities for a given dynamic matrix \({\bf M}\), referred to as the generalized Wick's theorem. Examples of the implementation of GF-free PTA are given in Sec.\ref{Sec6}. Finally, Sec.\ref{Sec7} provides a discussion and summary.

\end{section}

\begin{section}{Basic quantities obtained from a given dynamic matrix {\bf M}}
\label{Sec2}

In this section, we lay out the general structure of the GF-free formalism of PTA. 
For a given Hamiltonian $H$ that describes the system under study,
we select a set of basis operators $\{A_1, A_2, ..., A_D \}$ that are linearly independent. These basis operators span a $D$-dimensional subspace in the Liouville space. For example, for interacting fermion systems, according to the problem of interest and our knowledge about the excitations, we could choose the following types of operators to constitute the basis, single particle annihilation operators $c_{k}$, the particle-hole excitation operators $c_{k}^{\dagger}c_{k+q}$, the particle-particle excitation operators $c_{k}c_{-k+q}$, or one of the above types expanded by more complicated operators with the same quantum numbers.

The core assumption of PTA is that the Heisenberg EOM of the basis operators can be approximately written in the following form,
\begin{equation}    \label{Eq1}
   \left[ A_i, H \right] \approx \sum_{j=1}^{D} M_{ji} A_{j}.
\end{equation}
Here $\left[ X, Y \right] \equiv XY - YX$ is the commutator of operators $X$ and $Y$.  ${\bf M}$ is called the dynamic matrix. The eigen values of ${\bf M}$ are the excitation energies. In this regards, it is different from the dynamic matrix of phonon, whose eigen values are the square of the phonon energies. On the complete basis or the basis spanning a closed subspace for $\mathcal{L}$ (denoting its dimension $D_{exc}$), Eq.(\ref{Eq1}) is exact. The exact ${\bf M}$ is a matrix of full size $D_{exc} \times D_{exc}$. It depends only on the Hamiltonian parameters and is independent of the state information such as the temperature, electron filling, RDMs, etc. 

To split the above EOM into dynamical and static components, we note that for any operator $X$, we can split it into a sum $X = X_0 + X_d$ of the static component $X_0$ and the dynamical component $X_d$, with the respective definitions
\begin{eqnarray}   \label{Eq2}
&& X_0 = \sum_{m n (E_m = E_n)}  \langle m | X |n \rangle  |m \rangle \langle n |,  \nonumber \\
&& X_d = \sum_{m n (E_m \neq E_n)}  \langle m | X |n \rangle  |m \rangle \langle n |.  
\end{eqnarray}
Here, $|m\rangle$ is the eigen state of Hamliltonian $H$ with the eigen energy $E_m$, {\it i.e.}, $H |m \rangle = E_m | m \rangle$. The following properties hold for arbitrary operators $X$ and $Y$. (i) $X_0(t)=X_0(0)$; (ii) $[X_0, H]=0$; (iii) $\langle X_d \rangle =0$ and $\langle X_d Y_0 \rangle = \langle Y_0 X_d \rangle = 0$; (iv) $\langle X_0 Y_0 \rangle = \langle Y_0 X_0 \rangle$. Here, $X(t) = e^{iHt/\hbar} X e^{-iHt/\hbar}$ is the Heisenberg operator. $\langle ... \rangle$ denotes the average on the thermal equilibrium state of the Hamiltonian $H$ at temperature $T$. 
In general, the static component $X_0$ of a given operator $X$ is unknown.

 Because of $[X_0, H] = [X, H]_0 = 0$ for arbitrary operator $X$, we can split Eq.(1) into the equations for the dynamic and static components,
\begin{eqnarray}   \label{Eq3}
    &&  [ \vec{A}_{d}, H] \approx  {\bf M}^{T} \vec{A}_{d},   \nonumber \\
    &&   {\bf M}^{T} \vec{A}_{0} \approx 0.
\end{eqnarray}
Here, we have defined the column vectors $\vec{A}_d = (A_{1d}, A_{2d}, ..., A_{Dd})^{T}$ and $\vec{A}_0 = (A_{10}, A_{20}, ..., A_{D0})^{T}$.
For the basis that is complete or spans a closed subspace for the Liouville superoperator $\mathcal{L}$ (defined as $\mathcal{L} O = [O, H]$ for arbitrary operator $O$),  Eqs.(\ref{Eq1}) and (\ref{Eq3}) are exact. The space $\mathcal{S}$ spanned by $\{ A_i \}$ is the direct sum of the subspace $\mathcal{S}_d$ spanned by $\{ A_{id} \}$ and $\mathcal{S}_0$ spanned by $\{ A_{i0} \}$. That is, $\mathcal{S}= \mathcal{S}_d \oplus \mathcal{S}_0$. Suppose $\mathcal{S}_0$ and $\mathcal{S}_d$ have dimensions of $D_0$ and $D-D_0$, respectively. If $\{ A_i \}$ are linearly independent and $D_0 > 0$, both $\{ A_{id} \}$ and $\{ A_{i0} \}$ are linearly dependent bases. Then, ${\bf M}$ has $D_0$ zero eigenvalues. The corresponding $D_0$ eigen operators of $\mathcal{L}$ spans $\mathcal{S}_0$, the null space of $\mathcal{L}$.

For the incomplete basis, the above properties no longer hold exactly. In particular, $\mathcal{S} \neq \mathcal{S}_d \oplus \mathcal{S}_0$. Both $\{ A_{id} \}$ and $\{ A_{i0} \}$ are not necessarily linearly dependent bases. In that case, Eq.(\ref{Eq3}) may be contradicting and Eq.(\ref{Eq1}) will be problematic. ${\bf M}$ may have no zero eigenvalues even if $D_0 >0$. We call this problem about Eq.(\ref{Eq3}) and its consequence the static component problem of PTA. It's treatment will be discussed in the discussion section. For the moment, to derive the PTA equations, let us we assume that Eq.(\ref{Eq3}) holds without contradiction.

In the following, we will first assume that ${\bf M}$ in Eq.(\ref{Eq1}) is known. Starting from Eq.(\ref{Eq1}), we will derive the expressions for the basic quantities such as $\langle A_i^{\dagger} A_j \rangle$ and the spectral function $\rho_{A_i, A_j^{\dagger}}(\omega)$ in terms of ${\bf M}$. Higher order quantities of the basis operators can be calculated through the  generalized Wick's theorem, which will be presented in Sec.V. How to determine ${\bf M}$ will be discussed in Sec. IV. 

\begin{subsection}{ time evolution of $\vec{A}(t)$ and time correlation functions ${\bf C}(t-t^{\prime})$ and ${\bf D}(t-t^{\prime})$}

From Eq.(\ref{Eq1}), the Heisenberg equation of motion for the operator column vector $\vec{A}(t)$ can be written as $i\hbar \, d\vec{A}(t)/ dt \approx {\bf M}^{T} \vec{A}(t)$, which has the solution
\begin{equation}   \label{Eq3p}
    \vec{A}(t) = e^{\frac{1}{i\hbar} {\bf M}^{T} t}\vec{A}(0).
\end{equation}
Split into the dynamical and the static components, it gives $\vec{A}_d(t) = e^{\frac{1}{i\hbar} {\bf M}^{T} t}\vec{A}_d(0)$ and $\vec{A}_0(t)=\vec{A}_0(0)$. 
To guarantee the unitary time evolution of $\vec{A}_{d}(t)$, ${\bf M}$ must have only real eigen values and $D$ independent eigen vectors. 
We define the time correlation function matrices ${\bf C}(t, t^{\prime})$ and ${\bf D}(t, t^{\prime})$ as
\begin{eqnarray}    \label{Eq4p}
    && C_{ij}(t, t^{\prime}) \equiv \langle A_i^{\dagger}(t) A_{j}(t^{\prime}) \rangle,    \nonumber \\
   && D_{ij}(t, t^{\prime}) \equiv \langle A_i(t) A_{j}^{\dagger}(t^{\prime}) \rangle. 
\end{eqnarray}
They have the properties $[{\bf C}(t, t^{\prime})]^{\dagger}= {\bf C}(t^{\prime}, t)$ and $[{\bf D}(t, t^{\prime})]^{\dagger}= {\bf D}(t^{\prime}, t)$. They can be split into ${\bf C}(t, t^{\prime}) = {\bf C}_d(t, t^{\prime})+ {\bf C}_0$ and ${\bf D}(t, t^{\prime}) = {\bf D}_d(t, t^{\prime})+ {\bf D}_0$, with the definitions
\begin{eqnarray}    \label{Eq5p}
&& (C_d)_{ij}(t, t^{\prime}) \equiv \langle A_{id}^{\dagger}(t) A_{jd}(t^{\prime}) \rangle     \nonumber \\
&& (C_0)_{ij} \equiv \langle A_{i0}^{\dagger} A_{j0}\rangle,
\end{eqnarray}
and similarly for ${\bf D}_d(t, t^{\prime})$ and $ {\bf D}_0$.

Inserting Eq.(\ref{Eq3p}) into Eq.(\ref{Eq4p}), we obtain 
\begin{equation}   \label{Eq6p}
    {\bf C}(t, t^{\prime}) =  e^{-\frac{1}{i\hbar} {\bf M}^{\dagger} t} {\bf C}(0,0) e^{\frac{1}{i\hbar} {\bf M} t^{\prime}}.
\end{equation}
The time translation invariance of the equilibrium state of $H$ requires that ${\bf C}_d(t, t^{\prime}) ={\bf C}_d(t- t^{\prime})$. Denoting ${\bf C}(0,0)={\bf C}$, we obtain the relation
\begin{equation}   \label{Eq7p}
 {\bf C} {\bf M} = {\bf M}^{\dagger} {\bf C}  .
\end{equation}
Similarly, the time translation invariance ${\bf D}(t, t^{\prime})= {\bf D}(t-t^{\prime})$ requires that
\begin{equation}   \label{Eq8p}
 {\bf D}^{T} {\bf M} = {\bf M}^{\dagger} {\bf D}^{T},
\end{equation}
with $ {\bf D} = {\bf D}(0,0)$. Both ${\bf C}$ and ${\bf D}$ are Hermitian and positive semi-definite matrices. The above two equations, once split into the dynamical and static components as ${\bf C}={\bf C}_0 + {\bf C}_d$ and ${\bf D}={\bf D}_0 + {\bf D}_d$, give ${\bf C}_d {\bf M} = {\bf M}^{\dagger} {\bf C}_d$, ${\bf D}_d^{T} {\bf M} = {\bf M}^{\dagger} {\bf D}_d^{T}$, and ${\bf D}_0^{T} {\bf M} =  {\bf C}_0 {\bf M} = 0$.

Using Eqs.(\ref{Eq7p}) and (\ref{Eq8p}), one obtains the expression for the time correlation function of the basis operators as
\begin{equation}   \label{Eq9p}
    {\bf C}(t-t^{\prime}) = {\bf C} \, e^{-\frac{1}{i\hbar} {\bf M}(t-t^{\prime})}.
\end{equation}
Similarly, we have
\begin{equation}   \label{Eq10p}
    {\bf D}(t-t^{\prime}) = e^{\frac{1}{i\hbar} {\bf M^{T}}(t-t^{\prime})} \, {\bf D}.
\end{equation}

Eq.(\ref{Eq7p}) and (\ref{Eq8p}) can be extended. On any state described by the density operator $\rho$ satisfying $[\rho, H]=0$, the time correlation function $C[\rho]_{ij}(t-t^{\prime}) \equiv \text{Tr} [\rho A_{i}^{\dagger}(t) A_j(t^{\prime})]$ and $(D[\rho])_{ij}(t-t^{\prime}) \equiv \text{Tr} [\rho A_{i}(t) A_j^{\dagger}(t^{\prime})]$ have time translation invariance. Similar derivation gives the more general relations
\begin{equation}   \label{Eq11p}
 {\bf C}[\rho] {\bf M} = {\bf M}^{\dagger} {\bf C}[\rho], 
 \end{equation}
 and
 \begin{equation}   \label{Eq12p}
  {\bf D}[\rho]^{T} {\bf M} = {\bf M}^{\dagger} {\bf D}[\rho]^{T},
\end{equation}
with ${\bf C}_0[\rho] {\bf M} = {\bf D}_0[\rho]^{T} {\bf M} = 0$. 
Here, $\rho$ could be a thermal equilibrium state at an arbitrary temperature or an eigenstate of $H$. Equations (\ref{Eq7p}) and (\ref{Eq8p}) are the special case of Eqs.(\ref{Eq11p}) and (\ref{Eq12p}) for $\rho = e^{-\beta H}/{\text Tr}(e^{-\beta H})$.

Equations (\ref{Eq7p}) and (\ref{Eq8p}) are the basic physical constraints for ${\bf C}$ and ${\bf D}^{T}$ for a given ${\bf M}$. They also require that ${\bf M}$ be a quasi-Hermitian matrix. A matrix ${\bf X}$ is called quasi-Hermitian if there exists a Hermitian positive-definite matrix ${\bf G}$ such that ${\bf GX} ={\bf X}^{\dagger}{\bf G}$ \cite{scholtz1992quasihermitian}. A quasi-Hermitian matrix ${\bf X}$ has some notable properties. It can be diagonalized by the similarity transformation. All of its eigenvalues are real. That is, one can find an invertible matrix ${\bf U}$ such that ${\bf U}^{-1} {\bf X} {\bf U} = {\bf \Lambda}$ is a real diagonal matrix. At the same time, ${\bf U}$ can be chosen such that ${\bf U}^{\dagger} {\bf G} {\bf U} = \Lambda_{G} > 0$ is a positive diagonal matrix. Each diagonal element of ${\bf \Lambda}_{G}$ can be tuned by the arbitrary factor of the corresponding column of ${\bf U}$. 
For example, we can define a Hermitian matrix $\bf{X}_H = \bf{G}^{1/2}\bf{XG}^{-1/2}$ and find its unitary eigenvector matrix $\bf{U}_H$, with $\bf{U}_H^{-1}\bf{X}_H\bf{U}_H = \bf{\Lambda}$ being a real diagonal matrix. One can prove that $\bf{U} = \bf{G}^{-1/2}\bf{U}_H$ can diagonalize ${\bf X}$ as ${\bf U}^{-1} {\bf X} {\bf U} = {\bf \Lambda}$. For this specific ${\bf U}$, each of its columns has been normalized such that $\bf{U}^{\dagger}\bf{GU} = {\bf \Lambda}_G = \bf{1}$. 

Both ${\bf C}$ and ${\bf D}$ are Hermitian positive semi-definite matrices. Only for very special cases do they have zero eigenvalues. 
According to the above discussion, Eqs. (\ref{Eq7p}) and (\ref{Eq8p}) guarantee that there is an invertible transformation ${\bf U}$ such that ${\bf U}^{-1} {\bf M} {\bf U} = {\bf \Lambda} = \text{diag}(\lambda_1, \lambda_2, ..., \lambda_D)$ is a real diagonal matrix. This is consistent with the unitary evolution requirement for Eq.(\ref{Eq3p}). In this diagonal representation, Eq.(\ref{Eq1}) becomes $[O_{k}, H] \approx \lambda_k O_k$. The physical meaning is clear: $O_{k} = \sum_{i} U_{ik}A_{i}$ is the approximate eigen excitation operator with excitation energy $\lambda_k$. If the basis is complete, $O_k$
is either a dynamical operator $O_k = O_{kd}$ with $\lambda_k \neq 0$ or a conserved static operator $O_k = O_{k0}$ with $\lambda_k = 0$. If any $A_{i0}$ is nonzero, ${\bf M}$ will have at least one zero eigen value. Below, we suppose $\lambda_k = 0$ for $k \in [1, D_0]$ and $\lambda_k \neq 0$ for $k \in [D_0 + 1, D]$. That is,
\begin{equation}     \label{Eq12.5p}
     {\bf U}^{-1} {\bf M} {\bf U} = \left(
\begin{array}{cc}
{\bf 0} &  \bf{0}  \\
\bf{0} &   {\bf \Lambda}_d  \\
\end{array}
\right),
\end{equation}
with ${\bf \Lambda}_d$ being a $(D-D_0) \times (D-D_0)$ real diagonal matrix. For the incomplete basis, we expect that $[O_{k}, H] \approx 0$ for $k \in [1, D_0]$.

We can choose ${\bf U}$ such that it fulfills the generalized orthonormal relation ${\bf U}^{\dagger} {\bf C} {\bf U} = {\bf \Lambda}_C$, which means
${\bf U}^{\dagger} {\bf C} {\bf U} = \text{diag}(\langle O_{1}^{\dagger}O_1 \rangle, \langle O_{2}^{\dagger}O_2 \rangle, ..., \langle O_{D}^{\dagger}O_D \rangle )$. Similarly, ${\bf U}^{\dagger} {\bf D}^{T} {\bf U} = {\bf \Lambda}_{D} = \text{diag}(\langle O_{1}O_1^{\dagger} \rangle, \langle O_{2} O_2^{\dagger} \rangle, ..., \langle O_{D}O_D^{\dagger} \rangle )$. Eq.(\ref{Eq7}) to be presented below guarantees that ${\bf C}$ and ${\bf D}$ can be diagonalized by a common matrix ${\bf U}$. 
In the diagonal representation, the static and dynamical parts of ${\bf C}$ have the form
\begin{equation}     \label{Eq13p}
     {\bf U}^{\dagger} {\bf C} _0 {\bf U} = \left(
\begin{array}{cc}
({\bf \Lambda}_{C})_0 &  \bf{0}  \\
\bf{0} &   {\bf 0}  \\
\end{array}
\right)
\end{equation}
and
\begin{equation}     \label{Eq14p}
     {\bf U}^{\dagger} {\bf C}_d {\bf U} = \left(
\begin{array}{cc}
{\bf 0} &  \bf{0}  \\
\bf{0} &   ({\bf \Lambda}_{C})_d   \\
\end{array}
\right),
\end{equation}
where $({\bf \Lambda}_C)_0$ and $({\bf \Lambda}_C)_d$ are $D_0 \times D_0$ and
$(D-D_0 ) \times (D-D_0 )$ diagonal matrices, respectively.
${\bf D}^{T}$ has the same properties.

\end{subsection}

\begin{subsection}{Static Averages ${\bf C}$ and ${\bf D}$}


The basic static averages that we can directly calculate from PTA are the matrices ${\bf C}$ and ${\bf D}$. In the following, we show that Eq.({\ref{Eq1}}) can give a relation between ${\bf C}$ and ${\bf D}$ that will be useful for solving both of them. 
From Eq.(\ref{Eq1}), it is easy to prove the following identity for an arbitrary real number $\lambda$,
\begin{equation}    \label{Eq4}
  e^{-\lambda H} \vec{A} \,e^{\lambda H} \approx e^{\lambda {\bf M}^{T}} \vec{A} .
\end{equation}
Using the cyclic properties of the trace
\begin{equation}    \label{Eq5}
   C_{ij} = \frac{1}{Z} \text{Tr} \left[ A_{j} e^{-\beta H} A_{i}^{\dagger}  \right]  ,
\end{equation} 
we obtain
\begin{equation}    \label{Eq7}
    {\bf C} \, e^{\beta {\bf M}} \approx   {\bf D}^{T} .
\end{equation}
The time-dependent generalization of Eq.(\ref{Eq7}) can be obtained from Eqs.(\ref{Eq9p}) and (\ref{Eq10p}) as 
\begin{equation}    \label{Eq15}
       {\bf C}(t-t^{\prime}) \, e^{\beta {\bf M}} \approx [ {\bf D}(t^{\prime} - t)]^{T} .
\end{equation}
At $t=t^{\prime}$, this equation recovers Eq.(\ref{Eq7}). Essentially, this equation is the outcome of the fluctuation-dissipation theorem that connects the thermodynamics with the dynamics for an equilibrium state.

Note that the cyclic property of the trace Eq.(\ref{Eq5}) always holds for the grand canonical ensemble, but holds for the canonical ensemble only if $[H, N] = [A_i, N]=0$ ($i=1,2,..., D$). Therefore, Eqs.(\ref{Eq7}) and (\ref{Eq15}) apply only for these situations. Eq.(\ref{Eq7}) is compatible with the Hermiticity of ${\bf C}$ and ${\bf D}$, given Eqs.(\ref{Eq7p}) and (\ref{Eq8p}). Once split into the dynamical and the static components, Eq.(\ref{Eq7}) becomes $ {\bf C}_d \, e^{\beta {\bf M}} =   {\bf D}_d^{T}$ and $ {\bf C}_0  =   {\bf D}_0^{T}$. 
We can diagonalize $ {\bf C}$ and ${\bf D}^{T}$ using the same ${\bf  U}$ and produce $(\lambda_{C})_k e^{\beta \lambda_k} \approx (\lambda_{D})_{k}$ ($k \in [1, D]$). Here, $\lambda_k$, $(\lambda_{C})_k$ and $(\lambda_{D})_k$ are the eigenvalues of ${\bf M}$, ${\bf C}$, and ${\bf D}$, respectively.

Eq.(\ref{Eq7}) is the key relation of PTA that will be used to solve the static averages in the equilibrium state. Below we make some further discussions to it.
To facilitate the calculation, Eq.(\ref{Eq7}) can also be written in equivalent but different forms. For example, if the basis operators $\{ A_i \}$  are of boson nature, $\langle [A_{i}^{\dagger}, A_j] \rangle$ is a lower order average than ${\bf C}$ and ${\bf D}$. It is useful to define
\begin{eqnarray}    \label{Eq8}
    {\bf I}_{ij} \equiv \langle [A_{i}^{\dagger}, A_j] \rangle.
\end{eqnarray}
Considering ${\bf I} = {\bf C} - {\bf D}^{T}$ and ${\bf I}_0 = 0$, Eq.(\ref{Eq7}) can be written as 
\begin{equation}    \label{Eq9}
 {\bf C} = {\bf C}_0 - {\bf I} \left( e^{\beta {\bf M}} - \bf{1} \right)^{-1}.
\end{equation}
In case that ${\bf M}$ has zero eigenvalues, ${\bf I}$'s corresponding eigenvalues are also zero. The factor ${\bf I} \left( e^{\beta {\bf M}}- \bf{1} \right)^{-1}$ understood as the zero eigenvalues of ${\bf I}$ overriding the divergence of the eigenvalues of $\left( e^{\beta {\bf M}}- \bf{1} \right)^{-1}$. That is, 
\begin{eqnarray}      \label{Eq9.5}
&& {\bf I} \left( e^{\beta {\bf M}}- \bf{1} \right)^{-1}  \nonumber \\
&& = ({\bf U}^{\dagger})^{-1}  \left(
\begin{array}{cc}
{\bf 0} &  \bf{0}  \\
\bf{0} &    (\Lambda_{I})_{d}(e^{\beta \Lambda_d} -\bf{1}_d)^{-1}  \\
\end{array}
\right)  {\bf U}^{-1}. 
\end{eqnarray}
Here, $({\bf \Lambda}_{I})_d$, ${\bf \Lambda}_d$, and  ${\bf 1}_d$ represent the $D_0 \times D_0$ lower-right part of the diagonal matrices, which correspond to the dynamical components.

If $\{ A_i \}$  are of fermion nature, $\langle \{A_{i}^{\dagger}, A_j\} \rangle$ is a lower order average than ${\bf C}$ and ${\bf D}$. In that case, we can define
\begin{equation}    \label{Eq10}
    {\bf J}_{ij}  \equiv  \langle \{A_{i}^{\dagger}, A_j\} \rangle .
\end{equation}
Because of ${\bf J} = {\bf C} + {\bf D}^{T}$, we obtain
\begin{equation}    \label{Eq11}
 {\bf C} = {\bf J} \left( e^{\beta {\bf M}}+{\bf 1} \right)^{-1}.
\end{equation}
This form applies even if ${\bf M}$ is singular.

The self-consistent calculation of ${\bf C}$ and ${\bf I}$ (${\bf J}$) will be possible if ${\bf I}$ (${\bf J}$) can be expressed in terms of ${\bf C}$, i.e., if ${\bf I} = {\bf I}({\bf C})$ (${\bf J} = {\bf J}({\bf C})$) is explicitly known. In many practical problems, one can usually achieve this by selecting suitable basis operators $\{ A_i \}$, employing the algebraic relations of boson or fermion operators, or by introducing further decoupling approximations. 
If ${\bf M}$ is also expressed in terms of ${\bf C}$ and ${\bf I}$ (${\bf J}$), i.e., ${\bf M}={\bf M}({\bf C}, {\bf I})$ (${\bf M}({\bf C}, {\bf J})$), the complete self-consistent calculation of ${\bf C}$, ${\bf I}$ (${\bf J}$), and ${\bf M}$ need to be carried out. In the previous GF-based formalism of PTA \cite{fan2018projective,fan2019controllable,ma2021interacting,ma2022projective,fan2022spatial,jia2025thermal,wu2025self}, Eq.(\ref{Eq7}) was obtained using the spectral theorem of the GF.

At this point, it is beneficial to compare PTA with the variational RDM theory. As an incomplete description of the equilibrium state of $H$ at a given temperature $T$, Eq.(\ref{Eq7}) is obtained from the dynamical information Eq.(\ref{Eq1}). It is therefore not variational in the Feynman-Bogoliubov-Peierls sense and does not produce the upper bound of the true free energy. In the variational RDM theory \cite{mazziotti2007}, the ground state is estimated by varying its energy with respect to RDMs under certain constraints, such as the reducibility and N-representability of RDMs. Since usually these constraints are incomplete in the practical calculation, the resulting ground state energy is not the upper bound either. Both PTA and the variational RDM theories provide equations between static averages. In both theories, the algebraic properties of elementary boson or fermion operators and the algebraic constraints for the reduced density matrix elements are employed to facilitate the solution of the equations. 
Besides the information about static averages, PTA can also produce the dynamical information such as the excitation and spectral function.

\end{subsection}

\begin{subsection}{GF and spectral function}
\label{GFandSF}


Once we have obtained the self-consistent solution of ${\bf C}$ and ${\bf I}$ (${\bf J}$), we can proceed to produce the basic dynamical quantities from a one-shot calculation, such as the time correlation function ${\bf C}(t-t^{\prime})$ (Eq.(\ref{Eq9p})), ${\bf D}(t-t^{\prime})$ (Eq.(\ref{Eq10p})), GFs, and the spectral functions defined on $A_i$ operators. This shows the advantage of PTA compared to the existing state-targeting methods such as the variational RDM \cite{eugene2024variational} or the coupled cluster theory \cite{bartlett2007cc}. Although the time-dependent density matrix method \cite{shun1985explicit} can also produce dynamical quantities, the operators there are limited to density operators, which are of boson type only. In PTA, the basis operators $\{ A_i \}$ are general.

The boson-type retarded two-time GF for operators $A$ and $B$ is defined as
\begin{equation}    \label{Eq16}
G^{r}_{B}[A(t)|B(t^{\prime})] \equiv \frac{1}{i} \theta(t-t^{\prime}) \langle \left[A(t), B(t^{\prime}) \right]\rangle.
\end{equation}
Here, $X(t)=e^{iHt/\hbar} X e^{-iHt/\hbar}$ ($X=A,B$) is the Heisenberg operator. 
$\theta(t-t^{\prime})$ is the step function. 
$\langle ... \rangle$ denotes the average on the thermal equilibrium state of the Hamiltonian $H$ at temperature $T$. $\left[X,Y\right]=XY - YX$ is the commutator of $X$ and $Y$. 
The fermion-type GF for the operators $A$ and $B$ is defined with anti-commutator as
\begin{equation}    \label{Eq17p}
G^{r}_{F}[A(t)|B(t^{\prime})] \equiv \frac{1}{i} \theta(t-t^{\prime}) \langle \{A(t), B(t^{\prime}) \}\rangle.
\end{equation}
For both types of GF, the Fourier transformation reads
\begin{equation}   \label{Eq17}
   G^{r}(A|B)_{\omega} = \int_{-\infty}^{\infty}  G^{r}[A(t)|B(t^{\prime})] e^{i (\omega + i\eta)(t-t^{\prime}) } d(t-t^{\prime}).
\end{equation}
Here $\eta = 0^{+}$ is an infinitesimal positive number. 
In terms of the Zubarev GF $G(A|B)_{\omega}$ \cite{zubarev1960double} with $G(A|B)_{\omega+ i\eta} = G^{r}(A|B)_{\omega}$, the spectral function is defined as
\begin{equation}    \label{Eq19}
\rho_{A,B}(\omega) \equiv \frac{i}{2\pi} \left[ G(A|B)_{\omega+i\eta}  -  G(A|B)_{\omega -i\eta}\right].
\end{equation}
Note that the static component $A_0$ and $B_0$ do not contribute to $G^{r}_B(A|B)_{\omega}$ but contribute to $G^{r}_F(A|B)_{\omega}$.

We consider the GF matrix ${\bf G}_{ij}(\omega) \equiv G(A_i|A_{j}^{\dagger})_{\omega}$ and the corresponding spectral function matrix ${\bf \rho}_{ij}(\omega)$. The boson-type GF matrix can be write as ${\bf G}_{B}(t-t^{\prime}) = (1/i) \theta(t-t^{\prime})[{\bf D}(t-t^{\prime}) - {\bf C}^{T}(t^{\prime}-t) ]$. Inserting Eqs.(\ref{Eq9p}) and (\ref{Eq10p}) and carrying out the Fourier transformation, we obtain
\begin{equation}    \label{Eq20}
    {\bf G}_{B}(\omega) = - \left( \omega {\bf 1} - \frac{1}{\hbar}{\bf M}^{T} \right)^{-1} \, {\bf I}^{T}.
\end{equation}
The corresponding spectral function matrix is written as
\begin{equation}    \label{Eq21}
    {\bf \rho}_{B}(\omega) = -\delta\left(\omega {\bf 1} - \frac{1}{\hbar}{\bf M}^{T} \right) \, {\bf I}^{T}.
\end{equation}
Here, the delta function of a matrix $\delta({\bf X})$ is defined as $\delta( {\bf X}) \equiv {\bf U} \delta({\bf \Lambda}) {\bf U}^{-1}$ where ${\bf U}^{-1} {\bf X} {\bf U}= {\bf \Lambda}$ is the eigenvalue matrix of ${\bf X}$.
Note that both $ {\bf G}_{B}(\omega)$ and $ {\bf \rho}_{B}(\omega)$ have no static component, i.e.,
\begin{eqnarray}      \label{Eq22}
&& {\bf U}^{T}  {\bf G}_{B}(\omega) {\bf U}^{\ast}  = \left(
\begin{array}{cc}
{\bf 0} &  \bf{0}  \\
\bf{0} &    - ({\bf \Lambda}_{I})_{d} (\omega {\bf 1} - \frac{1}{\hbar}{\bf \Lambda}_d )^{-1}  \\
\end{array}
\right) .  \nonumber \\
&&
\end{eqnarray}

The quasi-Hermitian property of ${\bf M}$ guarantees that ${\bf \rho}_{B}(\omega)$ is well defined and ${\bf G}_{B}(\omega)$ has real first-order poles as it should. Therefore, the causality of the retarded GF and the analytical structure of GFs are fulfilled.
A similar derivation gives the fermion-type GF matrix and its spectral function as
\begin{equation}    \label{Eq23}
    {\bf G}_{F}(\omega) = \left( \omega {\bf 1} -\frac{1}{\hbar}{\bf M}^{T} \right)^{-1} \, {\bf J}^{T} 
\end{equation}
and
\begin{equation}    \label{Eq24}
    {\bf \rho}_{F}(\omega) = \delta\left(\omega {\bf 1} - \frac{1}{\hbar}{\bf M}^{T} \right) {\bf J}^{T} .
\end{equation}
Unlike ${\bf G}_{B}(\omega)$, the fermion-type GF has a pole at $\omega=0$ if $D_0 > 0$. We have
\begin{equation}    \label{Eq23n}
    {\bf G}_{F}(\omega) = \left( \omega {\bf 1} -\frac{1}{\hbar}{\bf M}^{T} \right)^{-1} \, {\bf J}_{d}^{T} +\frac{1}{\omega} \,{\bf J}_{0}^{T}.
\end{equation}
Correspondingly, the spectral function has a delta peak at $\omega =0$,
\begin{equation}    \label{Eq24n}
    {\bf \rho}_{F}(\omega) = \delta\left(\omega {\bf 1} - \frac{1}{\hbar}{\bf M}^{T} \right) {\bf J}_d^{T} + \delta(\omega) \, {\bf J}_{0}^{T}.
\end{equation}

The above expressions for the GF and spectral function have several desirable properties.

(i) If Eqs.(\ref{Eq7p}) and (\ref{Eq8p}) are satisfied, the PTA expressions for the Zubarev GF matrices Eqs.(\ref{Eq20}) and (\ref{Eq23}) satisfy the rigorous Hermitian symmetry
\begin{equation}    \label{Eq24p}
[{\bf G}_{B/F}(\omega)]^{\dagger} = {\bf G}_{B/F}(\omega) 
\end{equation}
for real $\omega$. 

(ii) The expressions for ${\bf \rho}_{B}(\omega)$ and ${\bf \rho}_{F}(\omega)$ in Eqs.(\ref{Eq21}) and (\ref{Eq24}) fulfill the exact relation between them, i.e.,
\begin{equation}   \label{Eq25}
  {\bf \rho}_F(\omega) \frac{1}{e^{ \beta \hbar\omega}+1}= {\bf \rho}_B(\omega) \frac{1}{e^{\beta \hbar\omega}-1} + {\bf C}_0^{T} \, \delta(\omega).
 \end{equation}

(iii) Using ${\bf J} {\bf M} = {\bf M}^{\dagger} {\bf J}$ and the fact that ${\bf J}$ is a positive semi-definite Hermitian matrix, one can prove that ${\bf \rho}_{F}(\omega)$ ( at arbitrary $\omega$) and ${\bf \rho}_{B}(\omega) \, \text{sgn}(\omega)$ (at $\omega \neq 0$) are positive semi-definite matrices. These are the properties of the exact spectral function matrix.

(iv) The exact spectral theorems for ${\bf \rho}_{B}(\omega)$ and ${\bf \rho}_{F}(\omega)$ 
\begin{equation}    \label{Eq26}
    \int_{-\infty}^{^{\infty}} d\omega \,{\bf \rho}_{B}(\omega) \frac{1}{e^{\beta \hbar \omega}-1} = {\bf C}_{d}^{T}
\end{equation}
and
\begin{equation}    \label{Eq27}
    \int_{-\infty}^{^{\infty}} d\omega \,{\bf \rho}_{F}(\omega) \frac{1}{e^{\beta \hbar \omega}+1} = {\bf C}^{T}
\end{equation}
are fulfilled. Indeed, inserting the expressions Eqs.(\ref{Eq21}) and (\ref{Eq24}) into the above equations, we recover the relation Eq.(\ref{Eq9}) and (\ref{Eq11}), respectively.

(v) Sum rule of the GFs.
The exact sum rules of the spectral function require that
\begin{equation}    \label{Eq28}
    \int_{-\infty}^{\infty} [{\bf \rho}_{B}(\omega)]_{ij} \, \omega^{n} d\omega = \langle [ ...[A_{i}, H], H,], ..., H], A_{j}^{\dagger}]\rangle
\end{equation}
and
\begin{equation}    \label{Eq29}
    \int_{-\infty}^{\infty} [{\bf \rho}_{F}(\omega)]_{ij} \, \omega^{n} d\omega = \langle \{ [...[A_{i}, H], H,], ..., H], A_{j}^{\dagger} \}\rangle.
\end{equation}
The right-hand side of the above equations contains $n$ commutators with $H$.
Inserting the PTA expression Eqs.(\ref{Eq21}) and (\ref{Eq24}) into the left side, we obtain
\begin{equation}    \label{Eq30}
    \left[ {\bf I} {\bf M}^{n}\right]_{ij} = \langle \left[ A_{i}^{\dagger}, [...[[A_j, H], H], ..., H] \right] \rangle
\end{equation}
and 
\begin{equation}    \label{Eq31}
    \left[ {\bf J}_d {\bf M}^{n}\right]_{ij} + \delta_{n,0} \,({\bf J}_0)_{ij}= \langle \left\{ A_{i}^{\dagger}, [...[[A_{j}, H], H], ..., H] \right\} \rangle,
\end{equation}
respectively.

For $n=0$, they give
\begin{eqnarray}    \label{Eq32}
    &&  {\bf I}_{ij} =  \langle [A_i^{\dagger}, A_j] \rangle, \nonumber \\
    &&  {\bf J}_{ij} =  \langle \{A_{i}^{\dagger}, A_{j} \} \rangle, 
\end{eqnarray}
which are nothing but the definitions of ${\bf I}$ and ${\bf J}$ matrices.
For $n=1$, Eqs.(\ref{Eq30}) and (\ref{Eq31}) give
\begin{eqnarray}    \label{Eq33}
    &&  ({\bf I M})_{ij} =  \langle [A_i^{\dagger}, [A_j, H]] \rangle \equiv ({\bf L}_{B})_{ij}, \nonumber \\
    &&  ({\bf J}_{d} {\bf M})_{ij} =  \langle \{A_{i}^{\dagger}, [A_{j}, H] \} \rangle \equiv ({\bf L}_{F})_{ij}.
\end{eqnarray}
They are the projection equation used in the previous PTA formalism to determine the dynamic matrix ${\bf M}$ self-consistently \cite{fan2018projective,fan2019controllable,fan2022spatial,ma2021interacting,wu2025self}. In this sense, the PTA equation conserves the $n=1$ sum rule even for a truncated operator basis. The matrices ${\bf L}_{B}$ and ${\bf L}_{F}$ are called Liouville matrices. Both are Hermitian matrices. The Hermiticity of them is related to the time-translation invariance of the equilibrium state via Eqs. (\ref{Eq7p}) and (\ref{Eq8p}), and is also related to the conservation identity Eq.(\ref{Eq37}). Note that in the practical calculation, preserving the Hermiticity of the Liouville matrix is an important non-trivial task \cite{fan2018projective, rowe1968eom,tohyama2004spurious}. ${\bf L}_{B}$ is negative semi-definite and ${\bf L}_F$ is indefinite. In previous PTA formalism, Eq.(\ref{Eq33}) is obtained by projecting Eq.(\ref{Eq1}) onto basis operators in order to obtain ${\bf M}$ self-consistently.
Here, in the present formalism, Eq.(\ref{Eq33}) is a requirement from the first-order sum rule of the spectral function.

\end{subsection}

\begin{subsection}{Transition matrix element and ground state energy}

In quantum chemistry and nuclear physics, one is often interested in the information about a specific excitation, such as the excitation energy and its transition matrix element. In condensed matter physics, such information is often integrated into the spectral function. Here, we show that PTA can give an expression similar to the one derived from Rowe's EOM method once the ${\bf C}$ and ${\bf D}^{T}$ are obtained.

Consider the ground state average $\langle O_{\nu}^{\dagger} O_{\nu} \rangle(T=0)$, with $O_{\nu}$ being the eigen excitation operator defined as $[O_{\nu}, H]=\lambda_{\nu }O_{\nu}$. We have $
\langle O_{\nu}^{\dagger} O_{\nu} \rangle(T=0) = \langle G | O_{\nu}^{\dagger}|\omega_{\nu} \rangle \langle \omega_{\nu} |O_{\nu} |G \rangle$, where $|\omega_{\nu} \rangle$ is the eigen state of $H$ with energy $E_G+ \omega_{\nu}$. Here we have assumed that $|\omega_{\nu} \rangle$ is nondegenerate. We obtain the transition amplitude for the excitation energy $\omega_{\nu}=-\lambda_{\nu} > 0$ as
\begin{equation}
   \langle \omega_{\nu} | O_{\nu} |G \rangle  = \Big\{ [ {\bf U}^{\dagger} {\bf C}(T=0) {\bf U}]_{\nu \nu} \Big\}^{1/2} e^{i \theta_{\nu}}.
\end{equation}
Similarly, from $\langle O_{\nu}O_{\nu}^{\dagger} \rangle = ({\bf U}^{\dagger}{\bf D}^{T}{\bf U} )_{\nu \nu}$ we obtain the transition amplitude for $\omega_{\nu}=\lambda_{\nu} > 0$ as
\begin{equation}
   \langle \omega_{\nu} | O_{\nu}^{\dagger} |G \rangle  = \Big\{ [ {\bf U}^{\dagger} {\bf D}^{T}(T=0) {\bf U}]_{\nu \nu} \Big\}^{1/2} e^{i \eta_{\nu}}.
\end{equation}
Here $\theta_{\nu}$ and $\eta_{\nu}$ are the phases that cannot be determined by the present method. Transforming back to the basis operator, we have
\begin{equation}
   \langle -\lambda_{\nu} | A_i |G \rangle  = ({\bf U}^{-1})_{\nu i}\Big\{ [ {\bf U}^{\dagger} {\bf C}(T=0) {\bf U}]_{\nu \nu} \Big\}^{1/2} e^{i \theta_{\nu}}
\end{equation}
and
\begin{equation}
   \langle \lambda_{\nu} | A_{i}^{\dagger} |G \rangle  = ({\bf U}^{-1})_{\nu i}^{\ast} \Big\{ [ {\bf U}^{\dagger} {\bf D}^{T}(T=0) {\bf U}]_{\nu \nu} \Big\}^{1/2} e^{i \eta_{\nu}}.
\end{equation}

The ground state energy is the key quantity to study in quantum chemistry and solid state physics. If we can write $H=\sum_{ij} h_{ij}A_i^{\dagger}A_j$, we have $E_G = \sum_{ij} h_{ij} {\bf C}_{ij}^{T=0}$. If the Hamiltonian can be expressed in the form $H = H_0 + \sum_{ij} h_{ij} A_i^{\dagger} A_j$, with $H_0$ being the non-interacting part whose ground state energy $E_{G0}$ is known exactly, we can calculate the ground state energy from the Feynman-Hellman theorem. We have
\begin{equation}
    E_G = E_{G0} +  \sum_{ij} h_{ij} \int_{0}^{1} dg \,{\bf C}_{ij}^{T=0}(g),
\end{equation}
where ${\bf C}^{T=0}(g)$ is the matrix ${\bf C}$ for the Hamiltonian $H= H_0 + g \sum_{ij} h_{ij} A_i^{\dagger} A_j$ at zero temperature.

\end{subsection}

\end{section}

\begin{section}{Physical Constraints of Averages}
\label{Sec3}
 
  In this section, we discuss possible physical constraints to the averages involved in ${\bf C}$ and ${\bf D}$ other than Eq.(\ref{Eq7}). Together with Eq.(\ref{Eq7}), they can be used to determine the solution for ${\bf C}$ and ${\bf D}$.

\begin{subsection}{algebraic constraints}

One type of the most frequently used constraints is the algebraic relations among the fermion or boson operators. Usually, ${\bf I}$ or ${\bf J}$ matrix contains the averages of lower order operators than ${\bf C}$ and ${\bf D}$, depending on the nature of the operators of $A_i$.
By selecting suitable basis operators, one could achieve such a situation that the explicit expressions for ${\bf I} = {\bf I}({\bf C})$ or ${\bf J} = {\bf J}({\bf C})$ can be written down. Then Eq.(\ref{Eq9}) or (\ref{Eq11})  becomes closed equations to solve (suppose ${{\bf C}_{0}}$ is known). For example, for interacting fermions, the operator basis $\{ a_{\alpha} \}$ composed of single-particle annihilation operators will produce ${\bf J}_{\alpha \beta} = \delta_{\alpha \beta}$. If we  assume that $({a}_{\alpha})_0 \approx 0$, we obtain ${\bf C} \approx (e^{\beta {\bf M} }+ {\bf 1})^{-1}$ from Eq.(\ref{Eq11}). 

\end{subsection}

\begin{subsection}{Hermiticity constraints}

There are a number of Hermiticity requirements for the matrices used in PTA. First of all, the matrices ${\bf C}$ and ${\bf D}$ are both Hermitian, with ${\bf C}^{\dagger} = {\bf C}$ and ${\bf D}^{\dagger} = {\bf D}$. 
For the given matrix ${\bf M}$, both ${\bf C}$ and ${\bf D}$ must satisfy the requirements Eqs.(\ref{Eq7p}) and (\ref{Eq8p}), which are equivalent to the Hermiticity of ${\bf C} {\bf M}$ and ${\bf D}^{T} {\bf M}$. They can be regarded as homogeneous linear equations for the elements of ${\bf C}$ and ${\bf D}$, respectively. Note that Eqs.(\ref{Eq7p}) and (\ref{Eq8p}), Eq.(\ref{Eq7}), and the Hermiticity of ${\bf C}$ and ${\bf D}$ are mutually compatible.
Other Hermiticity constraints include ${\bf L}_{B}^{\dagger} = {\bf L}_{B}$ and ${\bf L}_{F} ^{\dagger} = {\bf L}_{F}$, where ${\bf L}_{B}$ and ${\bf L}_{F}$ are the Liouville matrices defined in Eq.(\ref{Eq33}). Physically, they are related to the conserving properties to be discussed in Sec.III.D, but mathematically they are not independent from the two Hermiticity constraints discussed above.

\end{subsection}

\begin{subsection}{RDM constraints}
    
It is well known that the RDMs of a physical state have symmetries and constraints. The so-called N-representability condition of RDMs is a strong constraint on the averages. It has been one of the central topics in the study of variational RDM theory \cite{coleman1963structure,mazziotti2023quantum}. The averages of the form $\langle a_{\alpha}^{\dagger} a_{\beta}^{\dagger}\dots a_{\gamma} a_{\delta} \dots\rangle$ can be regarded as elements of a certain order of RDMs. Here, $a_{\alpha}^{\dagger}$ and $a_{\gamma}$, etc., are the creation and annihilation operators of a fermion or boson on the single particle orbitals $\alpha$ and $\gamma$, respectively.  One type of such constraints originates from the symmetry or anti-symmetry of exchanging two boson or fermion operators. For example, for fermions, one has
\begin{eqnarray}   \label{Eq34}
&& \langle a_{\alpha}^{\dagger} a_{\beta}^{\dagger} a_{\gamma} a_{\delta}\rangle \nonumber \\
&=& - \langle a_{\beta}^{\dagger} a_{\alpha}^{\dagger} a_{\gamma} a_{\delta}\rangle = - \langle a_{\alpha}^{\dagger} a_{\beta}^{\dagger} a_{\delta }a_{\gamma} \rangle  = \langle a_{\beta}^{\dagger} a_{\alpha}^{\dagger} a_{\delta} a_{\gamma}\rangle \nonumber \\ 
&=&  \delta_{\beta \gamma} \langle a_{\alpha}^{\dagger} a_{\delta} \rangle - \langle a_{\alpha}^{\dagger} a_{\gamma} a_{\beta}^{\dagger} a_{\delta}\rangle  = \delta_{\alpha \delta}  \langle a_{\beta}^{\dagger} a_{\gamma} \rangle - \langle a_{\beta}^{\dagger} a_{\delta} a_{\alpha}^{\dagger} a_{\gamma} \rangle. \nonumber \\
&&
\end{eqnarray}

 Another type of constraint of RDM comes from the reducibility, i.e., the partial trace of higher order RDM gives the lower RDM. One example is the relation used in the self-consistent RPA to produce the single particle density from the second order RDM \cite{wu2025self},
\begin{equation}    \label{Eq35}
   \langle a_{\alpha}^{\dagger}a_{\beta} \rangle = \frac{1}{L-N-1} \sum_{\gamma \neq \alpha, \beta} \langle (a_{\gamma}^{\dagger} a_{\alpha})^{\dagger} ( a_{\gamma}^{\dagger}a_{\beta} )\rangle \,\,\,\,(\alpha \neq \beta).
\end{equation}
Here, $N$ is the total number of fermions. $L$ is the number of single-particle orbitals. This equation is obtained from the number operator method \cite{rowe1968methods}. It is applicable only to the canonical ensemble. In the recent PTA-based self-consistent RPA studies, it is found that RDM-related constraints like Eqs.(\ref{Eq34}) and (\ref{Eq35}) significantly stabilize the iterative solution of the static averages \cite{wu2025self}. More N-representability conditions may also be considered.

\end{subsection}

\begin{subsection}{generalized virial theorem}
    
The generalized quantum virial theorem reads
\begin{equation}    \label{Eq36}
   \langle [A, B] \rangle = \langle e^{\beta H} [e^{-\beta H}, A] B \rangle = - \langle e^{\beta H} A [e^{-\beta H}, B] \rangle.
\end{equation}
Here, $A$ and $B$ are two arbitrary operators. This equation comes from the cyclic properties of the trace, and hence holds for the grand canonical ensemble as well as for the canonical ensemble when $[A, N]=[H, N]=0$.
Letting $B=H$, we get the conservation identity
\begin{equation}   \label{Eq37}
\langle [A, H] \rangle=0.
\end{equation}

Eq.(\ref{Eq36}) is called the generalized quantum virial theorem because one of the special cases of its classical counterpart is the virial theorem in the classical statistical physics. The classical version of Eq.(\ref{Eq36}) reads \cite{ma2022projective}
\begin{equation}    \label{Eq38}
  \langle \{ A, B\}_{P} \rangle = \beta \langle \{A, H\}_{P} B \rangle = - \beta \langle A \{B, H\}_{P} \rangle.
\end{equation}
Here, $A$ and $B$ are two classical dynamical variables. $H$ is the classical Hamiltonian. $\{.., ...\}_{P}$ is the Poisson bracket. For $B=H$, it gives
\begin{equation}    \label{Eq39}
    \langle \{ A, H \}_{P} \rangle = 0.
\end{equation}
For the special case $H= \sum_i \vec{p}_i^2/2m + V(\vec{r}_1, {\vec{r}_2, ..., {\vec{r}_N}})$ and $A = \sum_i \vec{r}_i \cdot \vec{p}_i$, this equation gives the virial theorem in the classical statistical mechanics $\sum_i \langle \vec{r}_i \cdot \nabla_i V \rangle = 2\sum_i \langle \vec{p}_i^2\rangle/2m$. Note that Eqs. (\ref{Eq36}) and (\ref{Eq37}) can be generalized from the ensemble average to the average on an eigenstate of $H$.

In PTA, The Hermiticity of ${\bf L}_B$ and ${\bf L}_F$ is a consequence of Eq.(\ref{Eq37}). But Eq.(\ref{Eq37}) can provide more identities that relate various static averages. They can be used as constraints for the unknown averages in ${\bf C}$, ${\bf D}$, or ${\bf M}$. Setting $A=A_i$, $A=A_i^{\dagger}A_j$, $A_i A_j A_k$, ..., etc., in Eq.(\ref{Eq37}) and using Eq.(\ref{Eq1}), one obtains a series of identities as
\begin{eqnarray}   \label{Eq39.5}
 && \sum_j M_{ji} \langle A_j\rangle = 0, \nonumber \\
 && \sum_k  \left[ M_{kj}\langle A_i^{\dagger}A_k\rangle - M_{ki}^{\ast} \langle A_k^{\dagger}A_j \rangle \right] = 0, \nonumber \\ 
 && \sum_l \left[ M_{li} \langle A_l A_j A_k \rangle + M_{lj} \langle A_i A_l A_k \rangle  + M_{lk} \langle A_i A_j A_l \rangle \right] = 0, \nonumber \\
 && ... 
\end{eqnarray}
These equations hold at the level of approximation of Eq.(\ref{Eq1}). The first line is consistent with the relation ${\bf M}^{T} \vec{A}_0 =0$ of Eq.(\ref{Eq3}) and the second line gives ${\bf CM}={\bf M}^{\dagger}{\bf C}$ of Eq.(\ref{Eq7p}). They are applicable to both the canonical and the grand canonical ensemble.

It is noted when applied to the density operators such as $a_{\alpha}^{\dagger}a_{\beta}$ and $a_{\alpha}^{\dagger}a_{\beta} a_{\gamma}^{\dagger} a_{\delta}$, Eq.(\ref{Eq37}) will produce the static relations that appears in the time-dependent density matrix theory \cite{shun1985explicit}.
For example, consider the Hamiltonian that contains only two-body interactions, 
\begin{equation}   \label{Eq37.1}
H=\sum_{\mu\nu}T_{\mu\nu}a_{\mu}^{\dagger}a_{\nu}+ \frac{1}{4}\sum_{\mu\nu\mu'\nu'}V_{\mu\nu\mu'\nu'}a_{\mu}^{\dagger}a_{\nu}^{\dagger}a_{\nu'}a_{\mu'},
\end{equation}
where the Hermiticity of $H$ requires that the coefficient matrices satisfy $T_{\mu\nu} = T_{\nu\mu}^{\ast}$ and $V_{\mu\nu\mu'\nu'} = V_{\mu'\nu'\mu\nu}^{\ast}$. This Hamiltonian is applicable to both fermions and bosons, with the symmetry constraint \(V_{\mu\nu\mu'\nu'}=\eta V_{\nu \mu \mu'\nu'} = \eta V_{\mu\nu\nu'\mu'} = V_{\nu\mu\nu'\mu'} \)  (\(\eta = -1\) for fermions or \(\eta = 1\) for bosons). Choosing 
$A=a_{\alpha}^{\dagger}a_{\beta}$, Eq.\eqref{Eq37} gives
\begin{align}  \label{Eq37.2}
 0=&\sum_{\mu}T_{\beta\mu}\langle a_{\alpha}^{\dagger}a_{\mu}\rangle +
\frac{1}{2}\sum_{\mu\mu'\nu'}V_{\beta\mu\mu'\nu'} \langle a_{\alpha}^{\dagger}a_{\mu}^{\dagger}a_{\nu'}a_{\mu'}\rangle\nonumber\\
 - &  \sum_{\mu}T_{\mu\alpha}\langle a_{\mu}^{\dagger}a_{\beta}\rangle - \frac{1}{2}
\sum_{\mu\mu'\nu'}V_{\mu'\nu'\alpha\mu} \langle a_{\mu'}^{\dagger}a_{\nu'}^{\dagger}a_{\mu}a_{\beta}\rangle  
\end{align}  
This equation establishes a constraint between the single-particle and two-particle RDMs.

\end{subsection}

\begin{subsection}{inequality constraints}

Besides the equality constraints discussed above, there are inequality constraints as well.
We have the following definiteness of matrices,
\begin{eqnarray}   \label{Eq40}
    &&  {\bf C} \succeq 0, \nonumber \\
    &&  {\bf D} \succeq 0, \nonumber \\
    && {\bf J} \succeq 0, \nonumber \\     
    && {\bf L}_{B} \preceq 0.
\end{eqnarray}
Here, ${\bf X} \succeq x$ ($\preceq$) means that the eigen values of the matrix ${\bf X}$ are larger (smaller) than or equal to $x$. Note that ${\bf I}$ and ${\bf L}_F$ are indefinite. 
At zero temperature and for a nondegenerate ground state, the Lehmann 
representation gives
\begin{equation}     \label{Eq40.5}
    ({\bf L}_B)_{ii} = - \sum_{n} \left(|\langle G | A_i^{\dagger}| n \rangle|^2 + |\langle n | A_i^{\dagger} | G \rangle|^2  \right) (E_{n} - E_{G} ).
\end{equation}
Here, $|n \rangle$ and $E_n$ are the eigen state and eige nenergy of $H$. $|G\rangle$ and $E_{G}$ are those for the ground state.
Therefore, the negativity of ${\bf L}_B$ at $T=0$ is related to the ground state stability. In the PTA-based RPA formalism, it is observed that the breaking down of negativity of ${\bf L}_B$ in the calculation is associated with the instability of the iterative solution \cite{wu2025self}. 

\end{subsection}

To summarize this section, we have construncted the following relations about the unknown averages ${\bf C}$ and ${\bf D}$ (for a given and fixed ${\bf M}$),
\begin{eqnarray}   \label{Eq41}
    && (1) \,\,\,\,\,\, {\bf C} \, e^{\beta {\bf M}} = {\bf D}^{T},  \nonumber \\
    && (2) \,\,\,\,\,\,  {\bf C} {\bf M} = {\bf M}^{\dagger} {\bf C}, \nonumber \\
    && (3) \,\,\,\,\,\, {\bf D}^{T} {\bf M} = {\bf M}^{\dagger} {\bf D}^{T}, \nonumber \\
    && (4) \,\,\,\,\,\, {\bf C}^{\dagger} = {\bf C},   \nonumber \\
    && (5) \,\,\,\,\,\, {\bf D}^{\dagger} = {\bf D},   \nonumber \\
    && (6) \,\,\,\,\,\, f_{\text{alg}}({\bf C},{\bf D}) = 0,    \nonumber\\
    && (7) \,\,\,\,\,\, {\bf C}\succeq 0,  \nonumber \\
   && (8) \,\,\,\,\,\, {\bf D} \succeq 0,  \nonumber \\
   && (9) \,\,\,\,\,\, {\bf L}_{B} \preceq 0
\end{eqnarray}
The equation $f_{\text{alg}}({\bf C},{\bf D}) = 0$ is an abbreviation for all the possible algebraic relations between ${\bf C}$ and ${\bf D}$, including the operator algebra constraints, the RDM constraints, and the identities from the generalized virial theorem. Since the other equations are homogeneous, $f_{\text{alg}}({\bf C},{\bf D}) = 0$ must contain the inhomogeneous relations necessary for a unique nonzero solution of ${\bf C}$ and ${\bf D}$. Note that the above relations are not independent. The independent relations are (1), (2), (4), (6), and (7).

The above equations are exact for a complete operator basis. Due to the quasi-Hermtian properties of ${\bf M}$, the matrices appearing in Eq.(\ref{Eq41}) can be diagonalized simultaneously.
That is, there exists an invertible matrix ${\bf U}$ (usually not a unitary matrix) such that
\begin{eqnarray}   \label{Eq42}
    && {\bf U}^{-1} {\bf M}{\bf U} = \Lambda_{M},  \nonumber \\
&& {\bf U}^{\dagger} {\bf C} {\bf U} = \Lambda_{C} \succeq  0,  
\nonumber \\    
&& {\bf U}^{\dagger} {\bf D}^{T} {\bf U} = \Lambda_{D}   \succeq 0,  
\nonumber \\    
&& {\bf U}^{\dagger} {\bf L}_B {\bf U} = \Lambda_{L_{B}} \preceq 0,   \nonumber \\
&& {\bf U}^{\dagger} {\bf L}_F {\bf U} = \Lambda_{L_{F}},    
\end{eqnarray}
are all real diagonal matrices. Eq.(\ref{Eq41}) can be transformed into the diagonal representation by ${\bf U}$.

Note that the above constraints may well be over complete and contradicting, unless the basis operators happen to span a closed subspace for $\mathcal{L}$. At the same time, the equations are nonlinear in the averages if ${\bf M}$ contains the unknown averages and needs to be determined self-consistently. This non-linearity could lead to multiple solutions. Therefore, in many practical calculations, one is faced with an over-constrained multiple-solution problem. Our practice with sc-RPA \cite{wu2025self} shows that the non-convergence in the strong coupling regime of the model is related to the fact that the averages cannot simultaneously satisfy the constraints that we exert. Therefore, instead of solving the PTA equations by conventional iterative scheme with all the constraints enforced by hand, we propose to seek for the optimal solution of the over-constrained problem Eq.(\ref{Eq41}). This issue will be further discussed in Sec.VII.B.

\end{section}

\begin{section}{Determination of ${\bf M}$}
\label{Sec4}


Up to now, our discussions are focused on the calculation of physical quantities for a given dynamic matrix ${\bf M}$. In this section, we discuss the possible ways to determine ${\bf M}$. 

On a complete basis $\{A_1, A_2,..., A_D \}$,  Eq.(\ref{Eq1}) becomes exact. In that case, the elements of ${\bf M}$ matrix, as the expansion coefficients of the commutators $[A_i, H]$, are determined solely by the Hamiltonian and the basis operators. Therefore, the exact ${\bf M}$ is independent of the state parameters such as temperature, electron filling, RDMs, etc. It also automatically satisfies all the constraints of ${\bf M}$ discussed above, such as its quasi-Hermitian properties Eqs.(\ref{Eq11p}) and (\ref{Eq12p}), and the sum rule of GFs Eqs.(\ref{Eq30})-(\ref{Eq31}).

In practical study, the basis usually does not span a closed subspace for ${\mathcal L}$. We have to find certain approximate expressions for ${\bf M}$. In previous as well as in the present PTA formalism, we confine our consideration to the projection method, in which we project Eq.(\ref{Eq1}) to $A_{k}$,
\begin{equation}   \label{Eq43}
    (A_{k} | [A_i, H]) = \sum_{j} M_{ji}(A_{k} | A_{j}).
\end{equation}
Here, $(X|Y)$ is the inner product between the two operators $X$ and $Y$. ${\bf M}$ is determined as the solution of the above linear equations once the inner products can be evaluated. This projective truncation of EOM gives the best approximation to $[A_i, H]$ in terms of the linear combination of $\{ A_{i} \}$ in the sense that the distance $|[A_i, H] - \sum_j M_{ji} A_{j}|$ measured by the given inner product is the smallest among all choices of ${\bf M}$. 

Note that the definition of the inner product is not unique. Mathematically, a product must satisfy the following requirements, 
\begin{eqnarray}   \label{Eq44}
&&    (X|Y) = (Y|X)^{\ast},  \nonumber \\
&&    (aX+bY|Z) = a^{\ast}(X|Y) + b^{\ast}(Y|Z),   \nonumber \\
&&    (Z|aX + bY) = a(Z|X) + b(Z|Y),   \nonumber \\    
&&    (X|X) > 0 \,\,\,\,\, \,\,\,\,  (\forall X \neq 0). 
\end{eqnarray}
 In the above equations, $X$ and $Y$ are arbitrary operators. $a$ and $b$ are arbitrary complex numbers. Eq.(\ref{Eq44}) is the definition for the positive definite inner product. In practice, some frequently used inner products are positive semi-definite, fulfilling Eq.(\ref{Eq44}) with the last line replaced by $(X|X) \geq 0$ ($\forall X \neq 0$). The indefinite inner product fulfills Eq.(\ref{Eq44}) without the last line. It defines the indefinite metric space \cite{hassi1994projection,Kawamura2006indefinite} and is also useful. For example, the commutator inner product $(X|Y)_B = \langle [X, Y] \rangle$ has been used to derive the self-consistent RPA formula within PTA. Although $(X|X)_B < 0$ occurs for certain $X \neq 0$, the obtained self-consistent RPA works pretty well in the weak to intermediate coupling regime \cite{wu2025self}.

We define the Liouville matrix as $L_{ij} \equiv (A_{i}|[A_j, H])$ and the inner product matrix as $K_{ij} \equiv (A_i|A_j)$. Eq.(\ref{Eq43}) becomes
\begin{equation}   \label{Eq45}
     {\bf L} = {\bf K} {\bf M}.
\end{equation}
To produce a quasi-Hermitian ${\bf M}$, both ${\bf L}$ and ${\bf K}$ must be Hermitian matrices and at least one of them is (positive- or negative-) definite.

If the inner product fulfills the last line of Eq.(\ref{Eq44}), ${\bf K}$ is positive definite and ${\bf M}$ is uniquely determined by Eq.(\ref{Eq45}) as ${\bf M}= {\bf K}^{-1} {\bf L}$. If the inner product does not fulfill the last line of Eq.(\ref{Eq44}), ${\bf K}$ could be singular or non-singular, depending on the choice of operator basis. For a singular ${\bf K}$, Eq.(\ref{Eq45}) provides a set of constraints that are insufficient to determine ${\bf M}$ uniquely.

\begin{table*}     \label{table1}
  \begin{center}
    \begin{tabular}{l|l|l} 
      \textbf{Name} &  \,  \textbf{Definition} & \textbf{Reference}\\
          \hline
      commutator inner product &  \,  $(X|Y)_B \equiv \langle [ X^{\dagger}, Y ] \rangle$  \, &  \cite{belkasri1994,winterfeldt1997,wu2025self} \\
      anti-commutator inner product &  \,  $(X|Y)_F \equiv \langle \{X^{\dagger}, Y\} \rangle$ \,  & \cite{roth1968new,onoda2001operator,kakehashi2004self,fan2018projective}\\
      double commutator inner product \, &  \,  $(X|Y)_D \equiv \langle [X^{\dagger}, [Y, H]] \rangle$  \,  & \cite{ma2022projective,jia2025thermal}\\
      Mori inner product & \,  $(X|Y)_M \equiv \frac{1}{\beta}\int_0^{\beta}d\tau\langle e^{\tau H}A^{\dagger}e^{-\tau H}B \rangle$  \,  & \cite{mori1965transport,mori1965continued,lee1982orthogonal}\\  
      infinite-$T$ inner product & \,  $(X|Y)_{\infty} \equiv \frac{1}{Z(\infty)} \text{Tr} (X^{\dagger}Y)$ \, &      \cite{florencio1987,viswanath1994recursion}\\        
    \end{tabular}
  \end{center}
   \caption{Some inner products used in the literature. Only a few representative references are listed here. The symbol $\langle ... \rangle$ represents the statistical average on the studied state and $Z(\infty)$ is the partition function at $T=\infty$. } 
\end{table*}

 For a complete basis, ${\bf M}$ is independent of the inner product. But for an incomplete basis, ${\bf M}$ depends on the inner product and so do the results of PTA. The inner product assigns the length to a given operator. Different inner products correspond to different ways of weighing the relative importance of the basis operators, and therefore influence the effectiveness of each of them. At the same time, the difficulty in evaluating the inner product varies significantly.

Various inner products have been used in the study of many-body systems since the idea of operator projection was developed in 1960's. We list some of them in Table 1 with the representative references. 
The commutator inner product $(X|Y)_B$ has been used in deriving the self-consistent RPA, where the excitation operators have boson character. The anti-commutator inner product $(X|Y)_F$ has been used in the study of single-particle excitations of fermion systems \cite{roth1968new,fan2018projective,fan2019controllable,ma2021interacting,fan2022spatial}. The double commutator inner product $(X|Y)_D$ is a quantum version of the one used in the PTA study of the classical systems \cite{ma2022projective,jia2025thermal}. The Mori inner product $(X|Y)_M$ has been used in deriving the Mori's formula \cite{mori1965transport,mori1965continued}. The infinite-$T$ inner product can be evaluated exactly and has been used in the study of dynamics of one-dimensional spin chain systems under Krylov basis \cite{florencio1987,viswanath1994recursion}, but it may not be the optimal one for describing low temperature properties. Note that at finite temperature, the listed inner products are positive definite except the commutator inner product. At zero temperature, except the commutator inner product and the infinite temperature inner product, all is positive semi-definite inner product. 

In the following, we consider two different ways of handling the inner product, the self-consistent evaluation that requires the input of ${\bf M}$ itself, and the non-self-consistent evaluation that is independent of ${\bf M}$. 

\begin{subsection}{self-consistent evaluation of inner product}

If the inner product $(X|Y)$ is defined as certain correlation function of $X$ and $Y$ on the equilibrium state of the studied system, its evaluation requires the PTA solution of the system and hence must be implemented self-consistently. Such inner products include the upper four examples in Table.1, with $\langle ... \rangle$ defined as the equilibrium state of $H$. 

To close the self-consistent equation regarding to ${\bf M}$, as done in the previous PTA implementations, we can express \( {\bf K } \) and \( {\bf L } \) in terms of ${\bf C }$ and ${\bf D}^{T}$, by selecting appropriate operator basis, using proper inner product, introducing additional approximations such as the partial projection approximation \cite{fan2018projective}, or solving additional EOM \cite{ma2022projective}. We can also use the generalized Wick's theorem presented in Sec.V for this purpose. Inserting the obtained ${\bf K}$ and ${\bf L}$ into Eq.(\ref{Eq45}), we obtain \( {\bf M } = {\bf M }({\bf C }, {\bf D }^{T}) \), which, together with Eq.\eqref{Eq9} or Eq.\eqref{Eq11}, forms the self-consistent equations for \( {\bf C} \), \( {\bf D}^{T} \), and \({\bf M}\).
Our practice of PTA shows that for an incomplete basis, the state-dependence of ${\bf M}$ from the self-consistency is quite important for PTA to give accurate results for the static as well as dynamical quantities at different temperatures. With increasing basis size, the self-consistently determined ${\bf M}$ will tend to the exact (state-independent) one.

A special class of the self-consistent inner product is the sum-rule-conserving inner product.
In Sec.\ref{GFandSF}, we show that if the inner product is consistent with the type of the used GF, PTA conserves the first-order sum rule of the spectral function Eq.(\ref{Eq33}). For example, to calculate the boson-type GF, we use $(X|Y)_B$. We then obtain ${\bf K} = {\bf I}$ and ${\bf L} = {\bf L}_B$. To calculate the fermion-type GF, $(X|Y)_F$ is used and we obtain ${\bf K} = {\bf J}$ and ${\bf L} = {\bf L}_F$. With these selections, Eq.(\ref{Eq45}) recovers the first-order sum rule constraint Eq.(\ref{Eq33}). This consistent choice of the inner product has some similarity in spirit to the spectral sum-rule method proposed by Nolting et al. \cite{geipel1988ferromagnetism,nolting1989band,nolting2009fundamentals}, where the fulfillment of sum rule is the guiding principle in constructing approximations.

For other arbitrary self-consistent inner product, although they will violate the first-order sum rule of certain type of the spectral function (boson-type or fermion-type), this violation is on the level of the PTA error and will be mitigated by enlarging the basis. For a decently large basis set, as the PTA error is small, it could be beneficial to use an inner product other than the GF-consistent one. 

We take RPA as an example. Traditional RPA can be realized by PTA on the basis of particle-hole excitation operators $A_{\alpha \beta} = c_{\alpha}^{\dagger} c_{\beta}$ with the commutator inner product. This inner product is consistent with the boson-type GF used in RPA \cite{wu2025self}. On the natural orbitals, the inner product matrix ${\bf I}$ has zero eigen value if the two single-particle orbitals have identical occupation, i.e., $\langle n_{\alpha}\rangle = \langle n_{\beta} \rangle$. This leads to difficulty to determine ${\bf M}$ from ${\bf L}_B = {\bf I}{\bf M}$. Thus, traditional RPA is not applicable to the high symmetry phases such as Mott insulator, spin liquid, or flat band system. On the contrary, the anti-commutator inner product has no such problem. It assigns a finite length to the excitation operator between degenerate orbitals. Therefore, there is strong motivation to seek for an improved RPA theory based on PTA with a different inner product. 

\end{subsection}

\begin{subsection}{non-self-consistent evaluation of inner product}
   
Under the self-consistent inner product discussed above, ${\bf M}$ needs to be solved self-consistently from Eq.(\ref{Eq7}), which now reads ${\bf C} e^{\beta \, {\bf M}({\bf C} , {\bf D}^{T} )} =   {\bf D}^{T}$. It is a  highly nonlinear equation and difficult to solve. Given the over-constraint nature of the problem as discussed in Sec.III, iterative solution often ends up with multiple solutions or no solution at all. The calculation of ${\bf L}$ matrix is also difficult if the inner product is to be evaluated self-consistently. This hampers the application of PTA to large basis set. To facilitate the implementation of PTA, a non-self-consistent inner product may be considered.

The infinite-$T$ inner product $(X|Y)_{\infty}$ in Table.1 is an example of such a non-self-consistent inner product. It is almost trivial to evaluate. But on an incomplete basis, it is difficult for this inner product to target the most interesting excitations, i.e., the low energy excitations from the ground state to the excited state. In order to remove this shortcoming, we may consider the Hartree-Fock inner product such as
\begin{equation}    \label{Eq46}
    (X|Y)_{B-HF} \equiv \langle [X^{\dagger}, Y]\rangle_{HF}.
\end{equation}
The inner product can also be defined on the equilibrium state of a soluble effective Hamiltonian, or the state described by a lower order approximation. On these states, both ${\bf K}$ and ${\bf L}$ are easy to evaluate and need to be evaluated only once. In this regards, Chatterjee and Pernal \cite{chatterjee2012excitation} used the variational density matrix theory to produce the matrices equivalent to ${\bf K}$ and ${\bf L}_B$ in the RPA equations, providing an example of the application of the non-self-consistent inner product in the RPA context. Our present work represents a general and systematic theory in this direction. 

In the complete basis limit, the ${\bf M}$ obtained from Eq.(\ref{Eq45}) is still exact. So are the static averages and the GF. It should be noted that in the expressions for GFs and spectral functions Eqs.(\ref{Eq20})-(\ref{Eq24}), ${\bf I}$ and ${\bf J}$ are the weight matrices determined self-consistently from ${\bf M}$, being different from the inner product matrix ${\bf K}$ in this case.
From the view of truncation of GF EOM, the averages introduced at the truncation are evaluated non-self-consistently, but those naturally appearing on the right-hand side of GF EOM are still determined self-consistently. 
The desirable properties of the GF and spectral functions (i) -(v) summarized in Eqs.(\ref{Eq24p})-(\ref{Eq33}) are independent of how ${\bf M}$ is determined. They still hold for PTA with non-self-consistent inner product, except that the first-order moment of the spectral function ${\bf \rho}_{B/F}(\omega)$ no longer agrees with the self-consistently determined ${\bf L}_{B/F}$.

Note that the PTA with Hartree-Fock inner product, when applied to the particle-hole excitation basis, generates a theory that is different from the standard RPA. In the standard RPA \cite{li2019}, not only ${\bf M}$ is evaluated on the Hartree-Fock ground state, but also the matrix ${\bf I}$ in the GF expression Eq.(\ref{Eq20}), leading to ${\bf I}={K}$ and the expression for GF as
\begin{equation}    \label{Eq47}
    {\bf G}_{B}(\omega) \approx - \left( \omega {\bf 1} - \frac{1}{\hbar}{\bf M}^{T} \right)^{-1} \, {\bf K}^{T},
\end{equation}
with $K_{ij} = \langle [A_i^{\dagger}, A_j]\rangle_{HF}$. This expression does not give the exact GF in the complete basis limit, since the correlations in the weight matrix ${\bf K}$ are neglected.

Finally, let us discuss what the inner product can influence and what it cannot. The definition of inner product will influence the Hermiticity of $\mathcal{L}$, but not the eigenvalues and eigen-operators of it. ${\bf M}$ is the linear representation of $\mathcal{L}$ in the operator basis. On the complete basis, the eigenvalues and eigen vectors of ${\bf M}$ faithfully agree with those of $\mathcal{L}$ and are independent of the inner product.

On an incomplete basis, $\mathcal{L}$ is represented in a subspace that is not closed. The properties of the representation matrix ${\bf M}$ now depends on the inner product. For the inner product on which $\mathcal{L}$ is an Hermitian operator, ${\bf M}$ still conserves the real eigen values (although ${\bf M}$ may not be a Hermitian matrix). But otherwise this is not guaranteed.
To respect the unitary time evolution of Heisenberg operators and the single real pole of GF, we require that ${\bf M}$ on the truncated operator subspace still have real eigen values. This can be achieved by enforcing the Hermiticity of ${\bf K}$ and ${\bf L}$ used in Eq.(\ref{Eq45}), i.e., by replacing ${\bf K}$ with $({\bf K} + {\bf K}^{\dagger})/2$ and ${\bf L}$ with $({\bf L} + {\bf L}^{\dagger})/2$. Then Eq.(\ref{Eq45}) can guarantee the real eigen value of ${\bf M}$. 

\end{subsection}

In summary of this section, unlike in previous PTA formalism, our present work formally separates the two issues of PTA, i.e., the calculation of physical quantities from a given ${\bf M}$, and the determination of ${\bf M}$. Thereby, it allows the choice of inner product that is independent of the GF's type. We hope that this work can lay the foundation for new theories of PTA with an optimal inner product. For the moment, exploration in this direction has not been fully pursued and awaits future research.

\end{section}

\begin{section}{Higher order quantities: Generalized Wick's Theorem}
\label{Sec5}


In this section, starting from Eq.(\ref{Eq1}), we show how the higher-order correlation functions of $A_i$, such as $\langle A_i(t_1) A_j(t_2) A_k(t_3) \rangle$, $ \langle A_{i}^{\dagger}(t_1)A_{j}^{\dagger}(t_2)A_{k}(t_3)A_{l} (t_4)\rangle$, and others can be systematically expressed in terms of \(\mathbf{M}\) and the lower-order averages. Such expressions can be used to close the self-consistent equations of PTA when necessary. They are also important for extending the scope of the theory and comparing theory with experiment. In the simplest basis $\{ a_{\alpha} \}$ of the single annihilation operators, the PTA expressions of higher order averages are exact for the non-interacting system and recover the Wick's theorem. In the complete basis limit, the obtained expressions are exact even for the interacting models. In general, with an incomplete basis, the obtained expressions are not exact for an interacting Hamiltonian. But as long as the the basis contains the operators that span a closed subspace for $\mathcal{L}_0 = [..., H_0]$, the obtained reduction relations are exact when the interaction tends to zero. Thus, they represent a generalization of the Wick's theorem to interacting Hamiltonians, giving a generalized factorization scheme of higher-order averages. We term the obtained formalism generalized Wick's theorem. Compared to the conventional Wick's theorem, the generalized Wick's theorem incorporates higher order correlations in a controlled manner when it is applied to the interacting systems. 

For any operator $B$, the average $\langle B \vec{A} \rangle$ satisfies 
\begin{equation}   \label{Eq3.1}
 \langle B\vec{A} \rangle = \langle \vec{A} (\beta)B \rangle,
\end{equation}
 where $\vec{A}(\beta) \equiv e^{\beta H} \vec{A} e^{-\beta H}$. Eq.\eqref{Eq3.1} holds for the grand canonical ensemble. It also applies to the canonical ensemble if \([\vec{A}, N] = [H, N] = 0\). Here, \(N\) is the operator for the total number of particles.
 Combining Eq.\eqref{Eq3.1} with Eq.\eqref{Eq4}, we obtain for a non-singular ${\bf M}$, 
\begin{align}   \label{Eq3.2}
&    \langle B\vec{A} \rangle \approx (e^{\beta {\bf M}^T}-{\bf 1})^{-1}\langle [\vec{A},B] \rangle,  \nonumber \\
&    \langle \vec{A} B \rangle \approx (e^{\beta {\bf M}^T}-{\bf 1})^{-1}e^{\beta {\bf M}^T}\langle [\vec{A},B] \rangle.
\end{align}
For a general ${\bf M}$, we also obtain
\begin{align}   \label{Eq3.3}
&    \langle B\vec{A} \rangle \approx (e^{\beta {\bf M}^T}+{\bf 1})^{-1}\langle \{\vec{A},B\} \rangle,  \nonumber \\
&    \langle \vec{A} B \rangle \approx (e^{\beta {\bf M}^T}+{\bf 1})^{-1}e^{\beta {\bf M}^T}\langle \{\vec{A},B\} \rangle.
\end{align}
The approximation signs in the above equations are due to that in Eq.(\ref{Eq1}).
The two sets of equations above are the foundation for our derivation of the generalized Wick’s theorem. 

When ${\bf M}$ has zero eigenvalues, Eq.(\ref{Eq3.2}) no longer applies. The correspondence of Eq.(\ref{Eq3.2}) is the more complicated equations
\begin{eqnarray}   \label{Eq3.3.1}
&&    \langle B \vec{A} \rangle = ({\bf U}^{T})^{-1}  \left(
\begin{array}{cc}
{\bf 1}_0 &  \bf{0}  \\
\bf{0} &   (e^{\beta {\bf \Lambda}_d} - {\bf 1}_d )^{-1}  \\
\end{array}
\right)
 \left(
\begin{array}{c}
{\bf U}_0^{T} \langle B \vec{A} \rangle  \\
{\bf U}_d^{T} \langle [\vec{A}, B] \rangle  \\
\end{array}
\right)   \nonumber \\
&& 
\end{eqnarray}
and 
\begin{eqnarray}    \label{Eq3.3.2}
&&   \langle \vec{A} B \rangle = ({\bf U}^{T})^{-1}  \left(
\begin{array}{cc}
{\bf 1}_0 &  \bf{0}  \\
\bf{0} &   (e^{\beta {\bf \Lambda}_d} - {\bf 1}_d )^{-1}e^{\beta {\bf \Lambda}_d}  \\
\end{array}
\right)
 \left(
\begin{array}{c}
{\bf U}_0^{T} \langle  \vec{A}B \rangle  \\
{\bf U}_d^{T} \langle [\vec{A}, B] \rangle  \\
\end{array}
\right) .  \nonumber \\
&&
\end{eqnarray} 
Here, $({\bf U}_0)_{D \times D_0}$ and $({\bf U}_d)_{D \times (D-D_0)}$ are the static and dynamical eigenvector matrices, respectively. The full ${\bf U}$ matrix is given by ${\bf U} = ({\bf U}_0, {\bf U}_d)$. In this case, Eqs.(\ref{Eq3.3.1}) and (\ref{Eq3.3.2}) can no longer realize the full order reduction. 

When \(\vec{A}\), \(B\), or both of them are the boson-type operators, Eq.\eqref{Eq3.2} realizes an order‑reduction by expressing higher-order quantities as combinations of lower-order ones. When both \(\vec{A} \) and \(B\) are the fermion-type operators, Eq.\eqref{Eq3.3} is required for the order reduction.
Here, a boson-type operator refers to the operator that is either a product of boson creation/annihilation operators, or a product of an even number of fermionic creation/annihilation operators, or an arbitrary linear combination of them. A fermion-type operator denotes the operator that is a product of an odd number of fermionic creation/annihilation operators, or arbitrary linear combination of them. 
It follows that the product, commutator, or anticommutator of two fermion-type operators is a boson-type operator.

Below, we will assume that ${\bf M}$ has no zero eigenvalues and employ Eqs.(\ref{Eq3.2}) and (\ref{Eq3.3}) to derive the generalized Wick's theorem. First, we consider higher-order correlation functions of the boson-type $\vec{A}$. We apply Eq.\eqref{Eq3.2} repeatedly. Each time we select a suitable $B$. Denoting the matrices ${\bf Q}_{-}= (e^{\beta {\bf M}^T}-{\bf 1})^{-1}$ and ${\bf P}_{-}= {\bf 1} + {\bf Q}_{-}$. we express the third-order correlation functions as
\begin{align}   \label{Eq3.4}
   & \langle A_i A_j A_k \rangle = {}\nonumber \\
   & \sum_{i_1,k_1}{\bf P}_{-,i i_1}{\bf Q}_{-,k k_1}\langle [A_{i_1}, [A_{k_1}, A_j]] \rangle \nonumber\\
 + & \sum_{j_1,k_1}{\bf Q}_{-,j j_1}{\bf Q}_{-,k k_1}\langle [A_{j_1}, [ A_{k_1}, A_i ] ] \rangle  . 
\end{align}
%
%
For the fourth-order correlation functions of \(\vec{A}\), we obtain
\begin{widetext}
\begin{align}
\label{Eq3.5}
& \langle A_i A_j A_k A_l \rangle
= {}  \nonumber\\
& \sum_{i_1,j_1,l_1}
{\bf P}_{-,i i_1}{\bf P}_{-,j j_1}{\bf Q}_{-,l l_1}
\left\langle
[ A_{j_1}, [A_{i_1}, [A_{l_1}, A_k ]]]
\right\rangle 
 + \sum_{i_1,l_1}
{\bf P}_{-,i i_1}{\bf Q}_{-,l l_1}
\left\langle
[A_{i_1} , A_j] [A_{l_1}, A_k ]
\right\rangle \nonumber\\
+ &  \sum_{i_1,k_1,l_1}
{\bf P}_{-,i i_1}{\bf Q}_{-,k k_1}{\bf Q}_{-,l l_1}
\left\langle
[A_{i_1}, [ A_{k_1}, [ A_{l_1}, A_j ]]]
\right\rangle 
 + \sum_{k_1,l_1}
{\bf Q}_{-,k k_1}{\bf Q}_{-,l l_1}
\left\langle
[ A_{k_1}, A_i] [ A_{l_1}, A_j]
\right\rangle \nonumber\\
+ &  \sum_{j_1,k_1,l_1}
{\bf Q}_{-,j j_1}{\bf Q}_{-,k k_1}{\bf Q}_{-,l l_1}
\left\langle
[ A_{j_1}, [ A_{k_1}, [A_{l_1}, A_i ]]]
\right\rangle 
+ \sum_{k_1,l_1}
{\bf Q}_{-,k k_1}{\bf Q}_{-,l l_1}
\left\langle
[A_{l_1}, A_i ] [A_{k_1}, A_j ]
\right\rangle .
\end{align}
\end{widetext}
The right-hand side involves only lower order quantities. Expressions for the averages involving $A_i^{\dagger}$ or both $A_i$ and $A_i^{\dagger}$ can be derived similarly. For examples, the reduction formulas for \(\langle A_i^{\dagger} A_j A_k  \rangle\), \(\langle A_i^{\dagger} A_j A_k A_l \rangle\), and \(\langle A_i^{\dagger} A_j^{\dagger} A_k A_l \rangle\) are presented in Appendix \ref{app:4-order quantity}. It is noted that the reduction formula for $\langle A_i A_j A_k \rangle$ and $\langle A_i A_j A_k A_l \rangle$ are not unique. By selecting different $B$ in applying Eq.(\ref{Eq3.2}), we can obtain formally different reduction formulas than Eqs.(\ref{Eq3.4}) and (\ref{Eq3.5}). They are exact for the complete basis, but differ at the level of PTA error for the incomplete basis. Note that to guarantee that the obtained expressions are exact in the non-interacting limit, we need to assure that every time we use Eq.(\ref{Eq3.2}) and (\ref{Eq3.3}), the involved $A_i$'s belong to the closed subspace for $\mathcal{L}_0$.

When $\vec{A}$ is of fermion-type,  Eq.\eqref{Eq3.5} does not yield an order reduction for the fourth-order quantities. Although Eq.\eqref{Eq3.4} provides a partial reduction, it is not optimal in the sense that the higher-order terms are not expressed in terms of the simplest possible lower-order quantities. In this case, we have to use Eqs.\eqref{Eq3.2} and \eqref{Eq3.3} alternatively, with appropriately selected operators $B$s. We obtain
\begin{align}   \label{Eq3.8}
   & \langle A_i A_j A_k  \rangle = {}\nonumber \\
   & \sum_{j_1,k_1}{\bf Q}_{-,j j_1}{\bf Q}_{-,k k_1}\langle [ A_{j_1} ,\{ A_{k_1} ,  A_i \}] \rangle \nonumber\\
   -  & \sum_{i_1,k_1}{\bf P}_{-,i i_1}{\bf Q}_{-,k k_1}\langle [ A_{i_1} , \{ A_{k_1} , A_j \}] \rangle .  
\end{align}
%
%
%
%
Denoting the matrices ${\bf Q}_{+}= (e^{\beta {\bf M}^T}+{\bf 1})^{-1}$ and ${\bf P}_{+}= {\bf 1} -{\bf Q}_{+} $, we obtain the expression for the forth order correlation function as
\begin{widetext}
\begin{align}
\label{Eq3.9}
& \langle A_i A_j A_k A_l  \rangle
= {}  \nonumber\\
& \sum_{j_1,k_1,l_1}
{\bf Q}_{+,j j_1}{\bf Q}_{+,k k_1}{\bf Q}_{+,l l_1}
\left\langle
\{ A_{j_1},
[ A_{k_1} ,\{  A_{l_1} , A_i \}]\}
\right\rangle 
+ \sum_{k_1,l_1}
{\bf Q}_{+,k k_1}{\bf Q}_{+,l l_1}
\left\langle
\{ A_{l_1} , A_i \}
\{ A_{k_1} , A_j \}
\right\rangle \nonumber\\
+ &  \sum_{i_1,k_1,l_1}
{\bf P}_{+,i i_1}{\bf Q}_{+,k k_1}{\bf Q}_{+,l l_1}
\left\langle
\{ A_{i_1} ,
[ A_{k_1} , \{ A_{l_1} ,  A_j \}]\}
\right\rangle  - \sum_{k_1,l_1}
{\bf Q}_{+,k k_1}{\bf Q}_{+,l l_1}
\left\langle
\{ A_{k_1} , A_i  \}
\{ A_{l_1} , A_j \}
\right\rangle \nonumber\\
- &  \sum_{i_1,j_1,l_1}
{\bf P}_{+,i i_1}{\bf P}_{+,j j_1}{\bf Q}_{+,l l_1}
\left\langle
\{ A_{j_1} ,
[ A_{i_1} , \{ A_{l_1} ,  A_k \}]\}
\right\rangle 
+ \sum_{i_1,l_1}
{\bf P}_{+,i i_1}{\bf Q}_{+,l l_1}
\left\langle
\{ A_{i_1} , A_j \}
\{ A_{l_1} , A_k \}
\right\rangle .
\end{align}
\end{widetext}
Similar reduction formulas for \(\langle  A_i ^{\dagger}  A_j A_k   \rangle\),
 \(\langle  A_i^{\dagger}  A_j A_k A_l  \rangle\) , and \(\langle  A_i^{\dagger} A_j^{\dagger}  A_k  A_l  \rangle\) with fermion-type $A_i$ are presented in Appendix \ref{app:4-order quantity}.

Eqs.\eqref{Eq3.2} and \eqref{Eq3.3} can be used to derive the order-reduction formula for the average of arbitrary product of $A_i$ and $A_i^{\dagger}$. The essential guiding principle in the derivation is as follows: For arbitrary operators \( X \) and \( Y \), we construct the commutator \([X, Y]\) when at least one of them is boson-type, and construct the anticommutator \(\{X, Y\}\) when both are fermion-type. For the time-dependent correlation functions of $A_i$ and $A_i^{\dagger}$ such as $\langle A_i(t_1) A_j(t_2) A_k(t_3) \rangle$, one can first use Eq.(\ref{Eq3p}) to reduce the time-dependent operators to the time-independent ones and then use Eqs.(\ref{Eq3.4})-(\ref{Eq3.9}). Therefore, higher-order multiple-time GF of $ \{ {A_i} \}$ and $\{ A_i^{\dagger} \}$ can be obtained. 

Note that the generalized Wick theorem presented here covers the previous method for calculating the higher-order average of the form $\langle B \vec{A} \rangle$. In previous works \cite{fan2018projective,ma2021interacting}, one usually write down the EOM for the GF $G(\vec{A}|B)_{\omega}$ and carry out the projective truncation. Here, with the general formalism, one is able to handle such issues in a unified manner. This will facilitate the derivation of ${\bf L}$ in terms of ${\bf M}$, ${\bf C}$, and ${\bf D}$ in the construction of the self-consistent equations. 

Although the traditional Wick's theorem only applies to the grand canonical ensemble, the present generalized Wick's theorem also applies to the canonical ensemble if $[A_i, N] = [H, N]=0$ is satisfied. The Wick's theorem generalized to the canonical ensemble \cite{ponomarenko2006,tsutsui2016,barghathi2020} is useful for the study of small finite systems such as molecules and cold atom systems. For such systems, the direct use of Wick's theorem will ignore the correlation due to the total particle number constraint and introduce errors. By using the RPA basis $\{ A_i = a_{\alpha}^{\dagger} a_{\beta} \}$ ($\alpha \neq \beta$), on which PTA is exact for the non-interacting limit, we obtain an alternative approach to the generalized Wick's theorem that is applicable for the canonical ensemble. The related work will be presented elsewhere.

\end{section}

\begin{section}{Examples}
\label{Sec6}


In this section, we apply the GF-free PTA formalism to some many-body Hamiltonians to demonstrate its application. 

\begin{subsection}{spinless fermions on a dimer}

We first consider an exact solvable two-site model of interacting spinless fermions, with the Hamiltonian
\begin{equation}    \label{Eq_examp1}
    H = \epsilon_1 n_1 + \epsilon_2 n_2 -t (a_1^{\dagger}a_2 + a_{2}^{\dagger}a_1) + V n_1 n_2.
\end{equation}
Here $n_1$ and $n_2$ are the fermion number operators on site $1$ and $2$, respectively. $a_1$ and $a_2$ are the annihilation operators of fermions. We assume that all the parameters are real numbers. In the case $t=0$, this model is equivalent to the model of the Hubbard atom. Here, we demonstrate the PTA equations and its solution at a special set of parameters $t^2 = \epsilon_1 \epsilon_2$, $V=-(\epsilon_1 + \epsilon_2)$, $\epsilon_1 > 0$, $\epsilon_2 > 0$, at which the basis operators have static components.

\begin{subsubsection}{Basis $\mathcal{B}_1 = \{ a_1, \, a_2, \, n_2 a_1, \, n_1 a_2 \}$ }

Writing down the commutator relations
\begin{eqnarray}   \label{Eq_examp2}
&&    [a_1, H] = \epsilon_1 a_1 - t a_2 + V n_2 a_1,  \nonumber \\
&&    [a_2, H] = \epsilon_2 a_2 - t a_1 + V n_1 a_2,  \nonumber \\
&&    [n_2 a_1, H] = (\epsilon_1 +V) n_2 a_1 - t n_1 a_2 , \nonumber \\
&&    [n_1 a_2, H] = (\epsilon_2 + V) n_1 a_2 - t n_2 a_1, 
\end{eqnarray}
we find  that this basis spans a closed subspace for $\mathcal{L}$. The exact dynamical matrix ${\bf M}$ in this subspace reads
\begin{equation}   \label{Eq_examp3}
    {\bf M} = \left(
\begin{array}{cccc}
\epsilon_1  &    -t        &   0           &   0   \\
-t          &  \epsilon_2  &   0           &   0   \\
V           &     0        & \epsilon_1+V  &  -t   \\
0           &     V        &   -t          & \epsilon_2+V   \\
\end{array}
\right).
\end{equation}
The eigen excitation operators (not normalized) and the corresponding excitation energies are obtained as
\begin{eqnarray}   \label{Eq_examp4}
&& O_1 = t a_1 + \epsilon_1 a_2 -t n_2 a_1 - \epsilon_1 n_1 a_2,     \,\,\,\, \,\,\,\, (\lambda_1 = 0)   \nonumber \\
&& O_2 = t a_1 - \epsilon_2 a_2 -t n_2 a_1 + \epsilon_2 n_1 a_2,     \,\,\,\, \,\,\,\, (\lambda_2 = -V)   \nonumber \\
&& O_3 = t n_2 a_1 + \epsilon_1 n_1 a_2,    \,\,\,\, \,\,\,\, (\lambda_3 = V)   \nonumber \\
&& O_4 = t n_2 a_1 - \epsilon_2 n_1 a_2.    \,\,\,\, \,\,\,\, (\lambda_4 = 0) 
\end{eqnarray}
$O_1$ and $O_4$ are conserved operators. This means that the basis operators contain static components. The self-consistent equation ${\bf C} e^{\beta {\bf M}}={\bf D}^{T}$ is written on the digonalized basis as $ \langle O_i^{\dagger}O_i \rangle e^{\beta \lambda_i} = \langle O_i O_i^{\dagger}\rangle$ ($i \in [1,4]$). They translate into the following equations for the averages $\langle n_1 \rangle$,  $\langle n_2 \rangle$, $\langle a_1^{\dagger}a_2 \rangle$, $ \langle a_2^{\dagger}a_1 \rangle$, and  $\langle n_1 n_2 \rangle$,
\begin{eqnarray}   \label{Eq_examp5}
&&  (\epsilon_1 + 2 \epsilon_2) \langle n_1 \rangle +  (\epsilon_2 + 2 \epsilon_1) \langle n_2 \rangle + t \langle p \rangle + 2V \langle n_1 n_2 \rangle \nonumber \\
&& = -V, \nonumber \\
&&  \nonumber \\
&& (Ve^{\beta V}- \epsilon_1  ) \langle n_1 \rangle +(Ve^{\beta V} - \epsilon_2 ) \langle n_2 \rangle + t \langle p \rangle 
\nonumber \\
&& - V(e^{\beta V}+1) \langle n_1 n_2 \rangle =  Ve^{\beta V},   \nonumber \\
&&  \nonumber \\
&&  \epsilon_1  \langle n_1 \rangle + \epsilon_2  \langle n_2 \rangle - t \langle p\rangle + V(e^{\beta V}+1) \langle n_1 n_2 \rangle = 0,   \nonumber \\
&&  \nonumber \\
&& \epsilon_2  \langle n_1 \rangle + \epsilon_1  \langle n_2 \rangle + t \langle p \rangle + 2V \langle n_1 n_2 \rangle = 0.
\end{eqnarray}
Here, $\langle p \rangle = \langle a_1^{\dagger}a_2 \rangle + \langle a_2^{\dagger}a_1 \rangle$. These equations apply only to the grand canonical ensemble. They are sufficient to give the four averages $\langle n_1 \rangle$, $\langle n_2 \rangle$, $\langle p \rangle$, and $\langle n_1 n_2 \rangle$. For the constraints provided by Eq.(\ref{Eq39.5}), ${\bf M}^{T} \langle \vec{A}\rangle = 0$ is trivial. ${\bf C}{\bf  M} = {\bf M}^{\dagger}{\bf C}$ gives
\begin{eqnarray}   \label{Eq_examp6}
   && \langle a_1^{\dagger}a_2 \rangle = \langle a_2^{\dagger}a_1 \rangle , \nonumber \\
   && t(\langle n_1 \rangle - \langle n_2 \rangle) = (\epsilon_2 - \epsilon_1) \langle a_2^{\dagger}a_1 \rangle,
\end{eqnarray}
which is applicable to both the canonical and the grand canonical ensembles. The first equation gives $\langle a_1^{\dagger}a_2 \rangle = \langle a_2^{\dagger}a_1 \rangle = \langle p \rangle/2$. The  second one is not independent and is consistent with Eq.(\ref{Eq_examp5}). The exact grand canonical averages are solved as $\langle n_1 \rangle = -[2\epsilon_2 + \epsilon_1 (1+e^{\beta V})]/(VZ)$, $\langle n_2 \rangle = 1 - \langle n_1 \rangle$, $\langle a_1^{\dagger}a_2 \rangle = \langle a_2^{\dagger}a_1 \rangle = -t(1-e^{\beta V})/(VZ)$, and $\langle n_1 n_2 \rangle = 1/Z$. Here $Z= 3 + e^{\beta V}$ is the grand partition function.

\end{subsubsection}

\begin{subsubsection}{Basis $\mathcal{B}_2 = \{m, \, h, \, p, \, n\}$ }

Basis $\mathcal{B}_2$ that we consider is composed of the particle-hole excitation operators used in RPA. Here, $m \equiv n_1 -n_2$, $h \equiv a_1^{\dagger}a_2 - a_2^{\dagger}a_1$, $p \equiv  a_1^{\dagger}a_2 + a_2^{\dagger}a_1$, and $n = n_1 + n_2$. On this basis, the dynamical matrix is 
\begin{equation}   \label{Eq_examp7}
    {\bf  M}  = \left(
\begin{array}{cccc}
0      &    -2t                  &          0                  &   0   \\
-2t    &     0                   &  \epsilon_2 -\epsilon_1     &   0   \\
0      & \epsilon_2 -\epsilon_1  &          0                  &  0   \\
0      &     0                   &          0                  &  0   \\
\end{array}
\right).
\end{equation}

The eigen excitation operators and the corresponding excitation energies read
\begin{eqnarray}   \label{Eq_examp8}
&& O_1 = -2t m + V h + (\epsilon_2 -\epsilon_1) p,     \,\,\,\, \,\,\,\, (\lambda_1 = V)   \nonumber \\
&& O_2 = -2t m - V h  + (\epsilon_2 -\epsilon_1) p,     \,\,\,\, \,\,\,\, (\lambda_2 = -V)   \nonumber \\
&& O_3 = (\epsilon_2 - \epsilon_1) m + 2tp,    \,\,\,\, \,\,\,\, (\lambda_3 = 0)   \nonumber \\
&& O_4 = n.    \,\,\,\, \,\,\,\, (\lambda_4 = 0) 
\end{eqnarray}
The operators $O_3$ and $O_4$ are conserved quantities. It turns out that the self-consistent equations $\langle O_i^{\dagger}O_i \rangle e^{\beta \lambda_i} = \langle O_i O_i^{\dagger} \rangle$ ($i \in [1,4]$) only give one independent equation (due to $O_1$), 
\begin{eqnarray}   \label{Eq_examp9}
 && \left[ V(\langle n \rangle - 2 \langle n_1 n_2 \rangle) -2t \langle p \rangle - (\epsilon_2 - \epsilon_1) \langle m \rangle \right]e^{\beta V}   \nonumber \\
 &=&   V(\langle n \rangle - 2 \langle n_1 n_2 \rangle) + 2t \langle p \rangle + (\epsilon_2 - \epsilon_1) \langle m \rangle .
\end{eqnarray}
The equation from $O_2$ is equivalent to this one since $O_2 = O_1^{\dagger}$. $O_3$ and $O_4$ gives trivial equations since they are Hermitian conserved quantities. This equation is applicable to both the canonical and the grand canonical ensembles. The conserving relations ${\bf M}^{T}\langle \vec{A} \rangle = 0$ and ${\bf CM} = {\bf M}^{\dagger}{\bf C}$ give out Eq.(\ref{Eq_examp6}). 
Therefore, for basis $\mathcal{B}_2$, PTA produces three independent equations for the five quantities. We can only solve them in the canonical ensemble where the number operator method can provide additional constraints for a fixed total fermion number $N = n_1+n_2$. This seems to be the general situation of RPA basis, for which one always needs to provide additional equations \cite{wu2025self}. Here, from $\langle a_1^{\dagger} N a_2 \rangle_{N} =  (N-1) \langle a_1^{\dagger}a_2 \rangle_N$, we obtain $\langle X \rangle_{N} = 0$ ($X= m, p, h, n$) for $N=0$ and $N=2$, with the exception $\langle n \rangle_{N=2} = 2$. For $N=1$, we  obtain the exact results $\langle m \rangle_{N} = [(\epsilon_2 - \epsilon_1)/V] \tanh{(\beta V/2)}$, $\langle p \rangle_{N} = (2t/V) \tanh{(\beta V/2)}$, $\langle h \rangle_{N} = \langle n_1 n_2 \rangle_{N} = 0$, and $\langle n \rangle_{N} = 1$.

\end{subsubsection}

\begin{subsubsection}{Basis $\mathcal{B}_3 = \{ a_1 a_2 \} $}

    This one-dimensional basis forms a closed subspace for $\mathcal{L}$ of the spinless dimer model. We have
\begin{equation}   \label{Eq_examp10}
    [a_1a_2, H] =  (\epsilon_1 + \epsilon_2 +V) a_1 a_2,
\end{equation}
which gives ${\bf M} = \epsilon_1 + \epsilon_2 +V $. The equation ${\bf C}e^{\beta {\bf M}} = {\bf D}^{T}$ produces 
\begin{equation}   \label{Eq_examp11}
   \langle n_1 n_2 \rangle e^{\beta (\epsilon_1 + \epsilon_2 +V )} = 1 - \langle n_1 \rangle - \langle n_2 \rangle + \langle n_1 n_2 \rangle.
\end{equation}
It applies only to the grand canonical ensemble. Equations ${\bf M}^{T} \langle \vec{A} \rangle = 0$ and ${\bf CM} = {\bf M}^{\dagger} {\bf C}$ are trivial. At the special parameter point $\epsilon_1 + \epsilon_2 = -V$, we obtain $O_1 = a_1a_2$, which is a conserved quantity with $\lambda_1 = 0$. Eq.(\ref{Eq_examp11}) is reduced to $\langle n_1 \rangle + \langle n_2 \rangle = 1$. This basis is another example of the situation where the exact dynamics in the closed operator subspace cannot produce sufficient thermodynamical equations.

\end{subsubsection}

\end{subsection}

\begin{subsection}{spinless fermions under Hartree-Fock basis}

Next, we apply PTA to the general quantum many-body Hamiltonian Eq.(\ref{Eq37.1}) for fermions, but take the single-particle annihilation operator basis, which we call Hartree-Fock (HF) basis. The Hamiltonian reads
\begin{equation}   \label{Eq37p}
H=\sum_{i j} T_{ij} a_{i}^{\dagger} a_{j}+ \frac{1}{4} \sum_{ij i' j'}V_{ij i' j'} a_{i}^{\dagger}a_{j}^{\dagger}a_{j'}a_{i'}.
\end{equation}
Here \(a_i^\dagger\) and \(a_i\) denote the single-particle creation and annihilation operators of fermions, respectively.  
The subscript \(i = 1, 2, ..., N\) is the single-particle index with the total number $N$. It could encompass all the indices such as spin, orbital, and so forth. $V_{iji'j'}$ is the anti-symmetrized interaction parameter, same as that in Eq.(\ref{Eq37.1}).

The HF basis is defined as $\mathcal{B}_{HF}= \{a_{1}, a_{2},..., a_{N} \}$. It is the simplest operator basis for describing the single-particle excitation of the system. It spans a subspace that is close for the Liouville superoperator $\mathcal{L}_0$ of the non-interacting Hamiltonian $H_0$. PTA on this basis gives the exact results for $V_{iji'j'}=0$.

For Hamiltonian Eq.(\ref{Eq37p}), we use the self-consistent anticommutator inner product and obtain the inner product matrix ${\bf K}$ and the Liouville matrix ${\bf L}$ as
\begin{equation}
    K_{ij} = \delta_{ij},
\end{equation}
and
\begin{equation}
    {\bf L}_{ij} = T_{ji} + \sum_{kl} V_{jkil} \langle a_{k}^{\dagger}a_{l} \rangle.
\end{equation}
The dynamic matrix is obtained as ${\bf M} = {\bf L}$, which is Hermitian. Note that for the HF basis, for the Hamiltonian that conserves the total particle number, ${\bf M}$ must be a Hermitian matrix, independent of the inner product selection. This can be proved by inserting Eq.(\ref{Eq1}) into the particle number conservation identity $[\sum_i A_i^{\dagger} A_i, H]=0$. 

The self-consistent equation Eq.(\ref{Eq11}) gives
\begin{equation}
    {\bf C} = \left( e^{\beta {\bf M}} + \bf{1}\right)^{-1} = {\bf Q}_{+}.
\end{equation}
The above equations are identical to the Hartree-Fock self-consistent equation.
The generalized Wick’s theorem formula derived in the preceding section gives
\(\langle a_i a_j a_k \rangle = \langle a_i^{\dagger} a_j a_k \rangle = \langle a_i a_j a_k a_l \rangle = \langle a_i^{\dagger} a_j a_k a_l \rangle = 0 \), and
\begin{eqnarray}   \label{Eq6.1}
   && \langle a_i^{\dagger} a_j \rangle = {\bf Q}_{+,ji},\nonumber \\
   &&  \langle a_i^{\dagger} a_j^{\dagger} a_k a_l \rangle = {\bf Q}_{+,li} {\bf Q}_{+,kj} - {\bf Q}_{+,ki} {\bf Q}_{+,lj} ,
\end{eqnarray}
which are same as those in the Hartree-Fock approximation. 
    
\end{subsection}

\begin{subsection}{spinless fermions under RPA basis}

As the last example of PTA application, we consider the same general many-body Hamiltonian for fermions Eq.(\ref{Eq37p}), but use the RPA basis which consists of the particle-hole operators $\mathcal{B}_{RPA} = \{ A_{(ij)} \equiv a_i^\dagger a_j \}$ ($i \neq j$). On this basis and with the self-consistent commutator inner product, PTA gives the self-consistent RPA \cite{wu2025self}. 
Note that the space spanned by $\mathcal{B}_{RPA}$ is closed with respect to the Hermitian conjugate operation. We have $(\vec{A}^{\dagger})^T={\bf S} ^T \vec{A} $, with \({\bf S}\) being a transformation matrix. The resulting symmetry of ${\bf M}$, ${\bf M} = -{\bf SM}^{\ast}{\bf S}^{-1} $, has many consequences in the results of the self-consistent RPA.
When other types of operators, such as $a_i a_j$, $a_{i}^{\dagger}a_{j}^{\dagger}$, and/or $a_{i}^{\dagger}a_{j}^{\dagger}a_{k}a_{l}$ are considered or added into $\mathcal{B}_{RPA}$, PTA will produce various (extended) versions of RPA \cite{hirsch2002fully,schuck2016progress}. 

For the basis $\mathcal{B}_{RPA}$ and with self-consistent commutator inner product, we obtain the inner product matrix {\bf K} as
\begin{equation}
   K_{kl,k'l'} = \delta_{k k'} \langle a_{k'}^{\dagger}a_{k} \rangle - \delta_{ll'} \langle a_l^{\dagger} a_{l'}\rangle.
\end{equation}
The Liuville matrix read
\begin{eqnarray}
  && L_{kl, k'l'}  \nonumber \\
  &=& \delta_{ll'} \sum_j T_{jk'} \langle a_{j}^{\dagger} a_k \rangle + \delta_{kk'} \sum_{j} T_{l' j} \langle a_l^{\dagger}a_j \rangle \nonumber \\
   && -T_{kk'} \langle a_{l}^{\dagger}a_{l'} \rangle - T_{l'l} \langle a_{k'}^{\dagger}a_{k} \rangle   \nonumber \\
   && + \frac{1}{2}\delta_{ll'} \sum_{i i'j'} V_{j' i' i k'} \langle a_{i'}^{\dagger} a_{j'}^{\dagger} a_i a_k\rangle  \nonumber \\
   && + \frac{1}{2} \delta_{kk'} \sum_{ii'j'} V_{l'ii'j'} \langle a_l^{\dagger}a_i^{\dagger}a_{j'}a_{i'} \rangle  \nonumber \\
&& -\frac{1}{2} \sum_{i' j'} \left[ V_{j' i' l k'} \langle a_{i'}^{\dagger} a_{j'}^{\dagger} a_{l'}a_{k}\rangle  + V_{l'ki'j'} \langle a_{l}^{\dagger}a_{k'}^{\dagger}a_{j'}a_{i'} \rangle \right]  \nonumber \\
&& - \sum_{ij} \left[ V_{jkik'} \langle a_l^{\dagger}a_{j}^{\dagger} a_i a_{l'} + V_{l' i l j} \langle a_{k'}^{\dagger} a_i^{\dagger} a_j a_k \rangle \rangle\right].
\end{eqnarray}
Suppose ${\bf C}_0=0$, the self-consistent equation (\ref{Eq9}) will be closed if it is supplemented with the following algebraic relations $\langle a_i^{\dagger}a_j^{\dagger}a_{k}a_l \rangle = \delta_{jk} \langle a_i^{\dagger}a_l\rangle - C_{ki, jl}$ and \cite{wu2025self}
\begin{eqnarray}
 && \langle a_{i}^{\dagger}a_j \rangle = \left\{\begin{array}{lll} 
 \frac{1}{N-N_e-1} \sum_{k \neq i,j} \langle (a_{k}^{\dagger} a_{i})^{\dagger} ( a_{k}^{\dagger}a_{j} )\rangle \,\,\,\, (i \neq j),  \nonumber \\
 \\
 \frac{1}{N-N_e} \sum_{k \neq i} \langle (a_{k}^{\dagger} a_{i})^{\dagger} ( a_{k}^{\dagger}a_{i} )\rangle  \,\,\, \,\,\,\,\,\, \,\,\, (i = j).  \nonumber 
 \end{array} \right. \\
 &&  
 \end{eqnarray}
Here, $N$ is the number of orbitals, and $N_e$ is the number of fermions. This expression is derived from the number operator method \cite{rowe1968methods} and applies only to the canonical ensemble.

For the RPA basis, the second-order quantities can be obtained from Eq.\eqref{Eq3.2} as
\begin{align}   \label{Eq6.2}
    & \langle A_{(ij)}A_{(kl)}\rangle = \langle a_i^{\dagger} a_j a_k^{\dagger} a_l \rangle \nonumber \\
     & = \sum_{i_1} {\bf Q}_{-,(kl)(i_1i)} \langle a_{i_1}^{\dagger} a_j\rangle - 
    \sum_{j_1} {\bf Q}_{-,(kl)(jj_1)} \langle a_{i}^{\dagger}a_{j_1}\rangle.
\end{align}
The generalized Wick's theorem gives the third-order quantity from Eq.\eqref{Eq3.4} as
\begin{align} \label{Eq6.3}
& & \langle A_{(ij)}A_{(kl)}A_{(mn)}\rangle = \langle a_i^{\dagger} a_j a_k^{\dagger} a_l c_m^{\dagger} c_n \rangle  \nonumber \\
 && = \sum_{i_1,j_1} P_{-,(ij)(i_1j_1)} Q_{-,(mn)(j_1k)} \langle a_{i_1}^{\dagger} a_l \rangle \nonumber \\
 && +  \sum_{i_1,j_1} P_{-,(ij)(i_1j_1)} Q_{-,(mn)(li_1)} \langle a_{k}^{\dagger} a_{j_1} \rangle \nonumber \\
 && + \sum_{i_1,j_1} Q_{-,(kl)(i_1j_1)} Q_{-,(mn)(j_1i)} \langle a_{i_1}^{\dagger} a_j \rangle \nonumber \\
 && +  \sum_{i_1,j_1} Q_{-,(kl)(i_1j_1)} Q_{-,(mn)(ji_1)} \langle a_{i}^{\dagger} a_{j_1} \rangle \nonumber \\
&& -  \sum_{i_1,j_1} R_{(ij)(kl)(mn),i_1j_1} \langle a_{i_1}^{\dagger} a_{j_1} \rangle,
\end{align}
with 
%
\begin{align}
    & R_{(ij)(kl)(mn),i_1j_1} = {} \nonumber \\
    & P_{-,(ij)(lj_1)} Q_{-,(mn)(i_1 k)} 
   + P_{-,(ij)(i_1 k)} Q_{-,(mn)(lj_1)} \nonumber \\
   +& Q_{-,(kl)(jj_1)} Q_{-,(mn)(i_1i)} 
   + Q_{-,(kl)(i_1i)} Q_{-,(mn)(jj_1)}  .   
\end{align}    
%

%
%
For the fourth-order quantities, Eq.\eqref{Eq3.5} gives
\begin{widetext}
\begin{align}   \label{Eq6.4}
   & \langle A_{(ij)}A_{(kl)}A_{(mn)}A_{(pq)}\rangle = \langle a_i^{\dagger} a_j a_k^{\dagger} a_l a_m^{\dagger} a_n a_p^{\dagger} a_q \rangle = {}   \sum_{i_1,j_1,k_1}S_{(ij)(kl)(mn)(pq),i_1j_1k_1}
    \langle a_{i_1}^{\dagger} a_{j_1} \rangle \nonumber \\
   & +  \sum_{i_1,j_1,k_1} P_{-,(ij)(i_1j_1)} P_{-,(kl)(k_1i_1)}Q_{-,(pq)(j_1m)} \langle a_{k_1}^{\dagger} a_n \rangle -
   \sum_{i_1,j_1,k_1} P_{-,(ij)(i_1j_1)} P_{-,(kl)(j_1k_1)}Q_{-,(pq)(ni_1)} \langle a_{m}^{\dagger} a_{k_1} \rangle \nonumber \\
   & + \sum_{i_1,j_1,k_1} P_{-,(ij)(i_1j_1)} Q_{-,(mn)(j_1k_1)}Q_{-,(pq)(k_1k)} \langle a_{i_1}^{\dagger} a_l \rangle -
   \sum_{i_1,j_1,k_1} P_{-,(ij)(i_1j_1)} Q_{-,(mn)(k_1i_1)}Q_{-,(pq)(lk_1)} \langle a_{k}^{\dagger} a_{j_1} \rangle \nonumber \\
   & + \sum_{i_1,j_1,k_1} Q_{-,(kl)(i_1j_1)} Q_{-,(mn)(j_1k_1)}Q_{-,(pq)(k_1i)} \langle a_{i_1}^{\dagger} a_j \rangle -
   \sum_{i_1,j_1,k_1} Q_{-,(kl)(i_1j_1)} Q_{-,(mn)(k_1i_1)}Q_{-,(pq)(jk_1)} \langle a_{i}^{\dagger} a_{j_1} \rangle   
\end{align}
\end{widetext}
\begin{widetext}
\begin{align}
\tag{\ref{Eq6.4} continued}
   & + \sum_{i_1,j_1} P_{-,(ij)(i_1k)} Q_{-,(pq)(j_1m)}
   \langle a_{i_1}^{\dagger} a_{l} a_{j_1}^{\dagger} a_{n} \rangle + 
   \sum_{i_1,j_1} P_{-,(ij)(li_1)} Q_{-,(pq)(nj_1)}
   \langle a_{k}^{\dagger} a_{i_1} a_{m}^{\dagger} a_{j_1} \rangle \nonumber \\
   & - \sum_{i_1,j_1} P_{-,(ij)(i_1k)} Q_{-,(pq)(nj_1)}
   \langle a_{i_1}^{\dagger} a_{l} a_{m}^{\dagger} a_{j_1} \rangle - 
   \sum_{i_1,j_1} P_{-,(ij)(li_1)} Q_{-,(pq)(j_1m)}
   \langle a_{k}^{\dagger} a_{i_1} a_{j_1}^{\dagger} a_{n} \rangle \nonumber \\
   & + \sum_{i_1,j_1} \big( Q_{-,(mn)(i_1i)} Q_{-,(pq)(j_1k)}
   + Q_{-,(mn)(j_1k)} Q_{-,(pq)(i_1i)} \big )
   \langle a_{i_1}^{\dagger} a_{j} a_{j_1}^{\dagger} a_{l} \rangle \nonumber \\
   & + \sum_{i_1,j_1} \big( Q_{-,(mn)(ji_1)} Q_{-,(pq)(lj_1)}
   + Q_{-,(mn)(lj_1)} Q_{-,(pq)(ji_1)} \big)
   \langle a_{i}^{\dagger} a_{i_1} a_{k}^{\dagger} a_{j_1} \rangle \nonumber \\
   & - \sum_{i_1,j_1} \big( Q_{-,(mn)(i_1i)} Q_{-,(pq)(lj_1)}
   + Q_{-,(mn)(lj_1)} Q_{-,(pq)(i_1i)} \big)
   \langle a_{i_1}^{\dagger} a_{j} a_{k}^{\dagger} a_{j_1} \rangle \nonumber \\
   & - \sum_{i_1,j_1} \big( Q_{-,(mn)(ji_1)} Q_{-,(pq)(j_1k)}
   + Q_{-,(mn)(j_1k)} Q_{-,(pq)(ji_1)} \big)
   \langle a_{i}^{\dagger} a_{i_1} a_{j_1}^{\dagger} a_{l} \rangle. \nonumber
\end{align}
\end{widetext}
The definition of \( S_{(ij)(kl)(mn)(pq),i_1j_1k_1} \) in Eq.\eqref{Eq6.4} is given in Appendix \ref{app:S}.
It should be noted that in Eqs.\eqref{Eq6.2}--\eqref{Eq6.4}, the notation \((ij)\) implicitly requires \(i \neq j\). 
Via Eq.\eqref{Eq3.2}, all \(\langle a^{\dagger}a a^{\dagger}a\rangle\)-type averages in Eq. \eqref{Eq6.4} can be further expressed as combinations of \(\langle a^{\dagger}a\rangle\)-type averages. As a result, both the third- and the fourth-order quantities of \(\vec{A}\) can be reduced to the \(\langle a^{\dagger}a\rangle\)-type averages.

\end{subsection}

\end{section}

\begin{section}{Discussion and Summary}
\label{Sec7}

In this section, we discuss two issues of PTA: The static component problem and the proposal to solve PTA equations as an optimization problem with constraints.

\begin{subsection}{static component problem}

 In this part, we discuss the static component problem of PTA. PTA is based on the ansatz Eq.(\ref{Eq1}), $[\vec{A}, H] \approx {\bf M}^{T} \vec{A}$. The selected basis operators $\{ A_i\}$ and ${\bf M}$ must fulfill the condition Eq.(\ref{Eq3}), i.e, $[\vec{A}_d, H] \approx {\bf M}^{T} \vec{A}_d$ and ${\bf M}^{T}\vec{A}_0 \approx 0$. We assume that the inner product is such that $\mathcal{S}_0$ and $\mathcal{S}_d$ are orthogonal. Then the projection of Eq.(\ref{Eq1}) to $A_k$ can be split into the projections of the above two equations to $A_{kd}$ and $A_{k0}$, respectively. We obtain
 \begin{eqnarray}    \label{Eq_disussion_1}
&&     (A_{kd}| [A_{id}, H]) \approx \sum_{j=1}^{D}{\bf M}_{ji} (A_{kd}|A_{jd}),   \nonumber \\
&&   \sum_{j=1}^{D}{\bf M}_{ji} (A_{k0}|A_{j0})   \approx 0.
 \end{eqnarray}
 In the case of complete basis, because of $\mathcal{S}=\mathcal{S}_0 \oplus \mathcal{S}_d$, the two equations are compatible. However, for an incomplete basis, the two equations may be contradicting. For example, if $\{ A_{id} \}$ contains $D$ linearly independent operators, ${\bf M}$ will be determined by the first equation uniquely. Then the second equation may not be fulfilled. The second equation means that if either the ranks of ${\bf M}$ or $\{ A_{i0} \}$, or both, must be linearly dependent. In practical calculations with a preselected incomplete basis $\{A_i \}$, more often than not, this requirement is not fulfilled.

 What is the consequence of results, if we calculate ${\bf M}$ from the projection of Eq.(\ref{Eq1}), i.e., from ${\bf L}={\bf K}{\bf M}$, knowing that some $A_{i0} \neq 0$ in the basis and Eq.(\ref{Eq1}) has a consistency problem? First, ${\bf L}={\bf K}{\bf M}$ amounts to the summation of the two equations of Eq.(\ref{Eq_disussion_1}). Since the second equation requires that ${\bf M}$ has zero eigenvalues while the first not, the resulting ${\bf  M}={\bf K}^{-1}{\bf L}$ is a compromise of the two requirements and will have finite eigenvalues $\lambda_{\nu} \neq 0$, with the corresponding excitation operator $O_{\nu}$ being mixed with static components. Then, the expected zero frequency excitations in the exact GF with $A_{i0} \neq 0$ emerge at small but finite frequencies. This is the spurious state problem in the formalism of Rowe's EOM \cite{rowe1968eom,chatterjee2012excitation}, due to violation of the killing condition $O_{\nu} |0\rangle =0$ for $\lambda_{\nu} > 0$ at $T=0$. Second, the problem will also increases errors in the PTA results since the finite static component on the right-hand side of Eq.(\ref{Eq1}) will distort ${\bf M}$ quantitatively by influencing the inner product matrix $(A_k|A_j)$.

 If the basis operators contain finite static components but they span a closed subspace or the full space for the Liouville operator $\mathcal{L}$, there will be no static component problem. As shown in the example of the spinless fermion dimer, by diagonalizing ${\bf M}$ and single out the zero eigen values, one can clearly separate the conserved operator $O_k$ ($k \in [1, D_0]$) with $\lambda_k =0$ from the dynamical excitations $O_k$ ($k \in [D_0+1, D]$) with $\lambda_k \neq 0$.  If the basis is incomplete but $A_i = A_{id}$ ($\forall i)$, there is no static component, either. For example, for \( A_i=[X_i,g(H)] \), one has \( A_{i0}=0 \). Here, \( X_i \) is an arbitrary operator and \( g(H) \) is an arbitrary function of the Hamiltonian. A concrete example is provided by the Krylov operator basis \( \vec{A}=(A_1, \mathcal{L}A_1, \mathcal{L}^2 A_1,\ldots)^T \) \cite{lee1982orthogonal,florencio1987,viswanath1994recursion}. Except for the first basis element, all Krylov basis operators have vanishing static components. In this case, ${\bf  M}$ does not have a zero eigenvalue. Therefore, we prefer to select the purely dynamical operators as our basis operators for the PTA calculation.

If the basis is incomplete and it does contain static component, $A_{i0} \neq 0$, there will be the static component problem. Below, we discuss several practical strategies to handle it. 
 If we cannot guarantee $A_{i0}=0$, we hope that their thermal expectations vanish, i.e., \( \langle A_{i}\rangle=0 \) ($\forall i$). Under this condition, one of the necessary condition ${\bf M}^{T} \langle\vec{A} \rangle = 0$ of the static component condition is satisfied without requiring that ${\bf M}$ has zero eigenvalues. Examples include: (i) pure fluctuation operators \( A_i=X_i-\langle X_i \rangle \); (ii) operators generated by conserved quantities, \( A_i=[X_i,g(Q)] \), where \( Q \) denotes any conserved charge associated with a exact symmetry of the Hamiltonian \( H \); and (iii) operators with finite quantum numbers, i.e., $A_i$ satisfying $[A_i, Q] = \delta q \, A_i$, with $Q$ being a conserved quantity of the system $[Q, H] = 0$, and $\delta q \neq 0$.  In particular, in quantum many-body systems obeying the eigenstate thermalization hypothesis, any local or quasi-local observable that breaks one or more exact symmetries is expected to have a vanishing static component \( A_{i0} = 0 \) in the thermodynamic limit \cite{srednicki1999approach,d2016quantum}.

Finally, we can recover the expected zero frequency poles in the GF with $A_{i0} \neq 0$ by modifying ${\bf M}$. If ${\bf U}^{-1} {\bf M} {\bf U} = \text{diag} ({\bf \Lambda}_0, {\bf \Lambda}_d )$, with the $D_0 \times D_0$ upper-left diagonal block ${\bf \Lambda}_0 \approx 0$, one can modify ${\bf M}$ into ${\bf M}_{\text{new}}$ as
\begin{equation}
   {\bf M}_{\text{new}} = {\bf U}  \left(
\begin{array}{cc}
{\bf 0}      &    {\bf 0}          \\
{\bf 0}      &    {\bf \Lambda}_d       \\
\end{array}
\right) {\bf U}^{-1}.
\end{equation}
Using the modified dynamical matrix ${\bf M}_{\text{new}}$ in the subsequent PTA calculation will recover $D_0$ zero frequency poles in the GF.

\end{subsection}

\begin{subsection}{PTA equation as an over-constraint optimization problem}
   
    The PTA self-consistent equation ${\bf C}e^{\beta {\bf M}}={\bf D}^{T}$ is solved usually with other constraints, as listed in Eq.(\ref{Eq41}). However, only when the basis is complete or spans a closed subspace for $\mathcal{L}$, can all the constraints be satisfied simultaneously. In the practical calculation with an incomplete basis, not all the constraints can be satisfied. This raises the concern that although some physical constraints are helpful for getting accurate solutions of Eq.(\ref{Eq41}), too much constraints may lead to over-constraint problem and hampers us from obtaining a solution at all.
    
    In the study of the one-dimensional spinless fermion model by the PTA-based sc-RPA method \cite{wu2025self}, we observed that without enforcing the algebraic constraints from the $N$-representability, the iterative solution of the sc-RPA equations only converges in the parameter regime $|V| \lesssim 0.07$. Enforcing some of the $N$-representability constraints, we can stablize the iterative solution to $V \lesssim 2.0$ or so, depending on the detailed iteration scheme. But for even larger interaction, the iterative solution does not converge. Apart from the factors of indefinite commutator inner product and the static component problem, we believe that the main reason for this is the over-constraint problem which somehow becomes more severe in the large interaction regime. Similar problem also appears in other form of RPA calculation \cite{rekkedal2013analytic,song2025unphysical}. 

One way to avoid this instability is to replace the iterative solution of PTA equation by solving an over-constraint optimization problem. In this scheme, the difference between the two sides of each PTA equation is minimized with respect to the RDMs, under all the possible constraints in Eq.(\ref{Eq41}). In this form, the solution of PTA equations is similar to that of the variational RDM theory \cite{mazziotti2004realization,mazziotti2007,eugene2024variational}.
The difference is that in the variational RDM theory, the quantity to be minimized is the ground state energy as a functional of the RDMs, while in PTA it is $|{\bf C}e^{\beta {\bf M}} - {\bf D}^{T}|$, which is derived from the truncated EOM of basis operators. The technique of semidefinite programming \cite{mazziotti2004realization} used in the variational RDM theory can be applied to PTA calculation. This perspective provides an interesting connection between the two theories.

\end{subsection}

  In summary, in this paper, we reformulate the PTA for GF EOM into a GF-free form. Although the obtained PTA equations for the static averages are essentially equivalent, the present formalism has a clearer mathematical structure, separating the two issues in PTA: Determining the dynamical matrix ${\bf M}$ and solving physical quantities from a given ${\bf  M}$. By doing so, the properties of 
 ${\bf M}$ is clarified and the inner product used to do the operator projection is detached from the type of the used GFs, opening wider possibility for constructing theories. The solution of the averages for a given ${\bf  M}$ is cast into an over-constraint optimization problem of RDMs to overcome the iteration instability problem. We discussed various sources of constraints for RDMs, including the generalized virial theorem. We also present the generalized Wick's theorem that can express the higher-order correlations in terms of the lower-order ones, which extends the application domain of PTA.

\end{section}
\begin{section}{Acknowledgments }
N.H.T. acknowledges helpful discussions with Y. Wan and previous collaborations on related topics with P. Fan, X.G. Ren, and H.W. Jia. This work is supported by National Natural Science Foundation of China (Grant Nos. 11974420).
\end{section}


\appendix{}

\begin{section}{
Other reduction formulas for several third- and fourth-order quantities}
\label{app:4-order quantity}
In this Appendix, we provide the reduction formulas for the 
third-order quantity \(\langle A_i^{\dagger} A_j A_k  \rangle\) and the
fourth-order quantities \(\langle A_i^{\dagger} A_j A_k A_l  \rangle\) and \(\langle A_i^{\dagger} A_j^{\dagger}  A_k A_l \rangle\) derived from Eqs. \eqref{Eq3.2} and \eqref{Eq3.3}. For the bosonic-type basis, we obtain 
\begin{align}
  & \langle A_i^{\dagger} A_j A_k \rangle ={}\nonumber \\
   & \sum_{i_1,j_1}{\bf Q}_{-,i i_1}^{\ast}{\bf P}_{-,j j_1}\langle [  A_{j_1}, [A_{k}, A_{i_1}^{\dagger} ]] \rangle  \nonumber\\
 + & \sum_{i_1,k_1}{\bf Q}_{-,i i_1}^{\ast}{\bf Q}_{-,k k_1}\langle [ A_{k_1} , [ A_{j} , A_{i_1}^{\dagger} ]] \rangle  ,
\end{align}
\begin{widetext}
\begin{align}   \label{Eq3.6}
 & \langle  A_i^{\dagger} A_j A_k A_l   \rangle = {}\nonumber\\
 & \sum_{i_1,j_1,k_1}{\bf Q}_{-,i i_1}^{\ast}{\bf P}_{-,j j_1}{\bf P}_{-,k k_1}\langle [ A_{k_1} , [(A_{j_1} , [ A_{l} ,  A_{i_1}^{\dagger} ]]] \rangle
 +  \sum_{i_1,j_1}{\bf Q}_{-,i i_1}^{\ast}{\bf P}_{-,j j_1}\langle [ A_{k}, A_{i_1}^{\dagger} ] [ A_{j_1} ,  A_l ] \rangle \nonumber\\
 + &  \sum_{i_1,j_1,l_1}{\bf Q}_{-,i i_1}^{\ast}{\bf P}_{-,j j_1}{\bf Q}_{-,l l_1}\langle [A_{l_1} ,[ A_{j_1} , [ A_{k} ,  A_{i_1}^{\dagger} ]]] \rangle 
 +  \sum_{i_1,j_1}{\bf Q}_{-,i i_1}^{\ast}{\bf P}_{-,j j_1}\langle [ A_{j_1} , A_{k} ] [ A_{l} , A_{i_1}^{\dagger} ] \rangle \nonumber\\
 + &  \sum_{i_1,k_1,l_1}{\bf Q}_{-,i i_1}^{\ast}{\bf Q}_{-,k k_1}{\bf Q}_{-,l l_1}\langle [A_{k_1} , [ A_{l_1} , [  A_{j} , A_{i_1}^{\dagger} ]]] \rangle   +  \sum_{i_1,l_1}{\bf Q}_{-,i i_1}^{\ast}{\bf Q}_{-,l l_1}\langle [ A_{j} , A_{i_1}^{\dagger} ] [ A_{l_1} , A_k ] \rangle   ,  
\end{align}
 and
\begin{align}   \label{Eq3.7}
 & \langle A_i^{\dagger} A_j^{\dagger} A_k A_l \rangle = {}\nonumber\\
 & \sum_{i_1,j_1,k_1}{\bf Q}_{-,i i_1}^{\ast}{\bf Q}_{-,j j_1}^{\ast}{\bf P}_{-,k k_1}\langle [ A_{k_1}, [[ A_{l}, A_{i_1}^{\dagger}], A_{j_1}^{\dagger} ]] \rangle
 +  \sum_{i_1,j_1}{\bf Q}_{-,i i_1}^{\ast}{\bf Q}_{-,j j_1}^{\ast}\langle [ A_{k}, A_{j_1}^{\dagger} ] [ A_{l}, A_{i_1}^{\dagger} ] \rangle  \nonumber\\
 + & \sum_{i_1,j_1,l_1}{\bf Q}_{-,i i_1}^{\ast}{\bf Q}_{-,j j_1}^{\ast}{\bf Q}_{-,l l_1}\langle [A_{l_1}, [[ A_{k}, A_{i_1}^{\dagger} ], A_{j_1}^{\dagger} ]] \rangle 
 +  \sum_{i_1,j_1}{\bf Q}_{-,i i_1}^{\ast}{\bf Q}_{-,j j_1}^{\ast}\langle [ A_{k}, A_{i_1}^{\dagger} ] [A_l, A_{j_1}^{\dagger}] \rangle \nonumber\\
 + & \sum_{i_1,k_1,l_1}{\bf Q}_{-,i i_1}^{\ast}{\bf Q}_{-,k k_1}{\bf Q}_{-,l l_1}\langle [ A_{k_1}, [A_{l_1}, [ A_{j}^{\dagger}, A_{i_1}^{\dagger} ]]] \rangle  
 +  \sum_{i_1,l_1}{\bf Q}_{-,i i_1}^{\ast}{\bf Q}_{-,l l_1}\langle [ A_{j}^{\dagger}, A_{i_1}^{\dagger} ] [ A_{l_1}, A_k] \rangle  .
\end{align}
%
For the fermionic-type basis operators, we find 
\begin{align}
     & \langle A_i^{\dagger} A_j A_k \rangle = {} 
    \sum_{j_1,k_1}{\bf Q}_{-,j j_1}{\bf Q}_{-,k k_1}\langle [ A_{j_1}, \{A_{k_1}, A_{i}^{\dagger} \}] \rangle  
   -  \sum_{i_1,k_1}{\bf Q}_{-,i i_1}^{\ast}{\bf Q}_{-,k k_1}\langle [\{ A_{k_1}, A_{j} \}, A_{i_1}^{\dagger} ] \rangle,   
\end{align}
%
%
\begin{align} \label{Eq3.10}
& \langle A_i^{\dagger} A_j A_k A_l \rangle = {}\nonumber\\
& \sum_{j_1,k_1,l_1}{\bf Q}_{+,j j_1}{\bf Q}_{+,k k_1}{\bf Q}_{+,l l_1}\langle \{ A_{j_1}, [ A_{k_1}, \{ A_{l_1}, A_i^{\dagger} \}] \}\rangle 
  + \sum_{k_1,l_1}{\bf Q}_{+,k k_1}{\bf Q}_{+,l l_1}\langle \{ A_{l_1}, A_{i}^{\dagger} \} \{ A_{k_1}, A_j \} \rangle  \nonumber\\
 + & \sum_{i_1,k_1,l_1}{\bf Q}_{+,i i_1}^{\ast}{\bf Q}_{+,k k_1}{\bf Q}_{+,l l_1}\langle \{ A_{i_1}^{\dagger}, [ A_{k_1}, \{ A_{l_1}, A_j \}] \} \rangle  
  -  \sum_{k_1,l_1}{\bf Q}_{+,k k_1}{\bf Q}_{+,l l_1}\langle \{ A_{k_1}, A_{i}^{\dagger} \} \{A_{l_1}, A_j \} \rangle \nonumber \\
 + & \sum_{i_1,j_1,l_1}{\bf Q}_{+,i i_1}^{\ast}{\bf P}_{+,j j_1}{\bf Q}_{+,l l_1}\langle \{ A_{j_1}, [\{ A_{l_1}, A_{k} \}, A_{i_1}^{\dagger} ] \} \rangle  +  \sum_{i_1,l_1}{\bf Q}_{+,i i_1}^{\ast}{\bf Q}_{+,l l_1}\langle \{ A_{j}, A_{i_1}^{\dagger} \} \{ A_{l_1}, A_k \} \rangle  ,
\end{align}
 and
\begin{align}
\label{Eq3.11}
& \langle A_i^{\dagger} A_j^{\dagger} A_k A_l \rangle
= {}  \nonumber\\
& \sum_{j_1,k_1,l_1}
{\bf P}_{+,j j_1}^{\ast}{\bf Q}_{+,k k_1}{\bf Q}_{+,l l_1}
\langle
\{ A_{j_1}^{\dagger},
[A_{k_1}, \{ A_{l_1}, A_i^{\dagger} \}]\}
\rangle
+ \sum_{k_1,l_1}
{\bf Q}_{+,k k_1}{\bf Q}_{+,l l_1}
\langle
\{ A_{l_1}, A_i^{\dagger} \}
\{ A_{k_1}, A_j^{\dagger} \}
\rangle \nonumber\\
 + & \sum_{i_1,j_1,l_1}
{\bf Q}_{+,i i_1}^{\ast}{\bf Q}_{+,j j_1}^{\ast}{\bf Q}_{+,l l_1}
\langle
\{ A_{j_1}^{\dagger},
[\{ A_{l_1}, A_k\}, A_{i_1}^{\dagger} ]\}
\rangle
+ \sum_{i_1,l_1}
{\bf Q}_{+,i i_1}^{\ast}{\bf Q}_{+,l l_1}
\langle
\{ A_j^{\dagger}, A_{i_1}^{\dagger} \}
\{ A_{l_1}, A_k\}
\rangle \nonumber\\
 + & \sum_{i_1,k_1,l_1}
{\bf Q}_{+,i i_1}^{\ast}{\bf Q}_{+,k k_1}{\bf Q}_{+,l l_1}
\langle
\{ A_{i_1}^{\dagger},
[ A_{k_1}, \{ A_{l_1}, A_j^{\dagger} \}] \}
\rangle
- \sum_{k_1,l_1}
{\bf Q}_{+,k k_1}{\bf Q}_{+,l l_1}
\langle
\{ A_{k_1}, A_i^{\dagger} \}
\{ A_{l_1}, A_j^{\dagger}\}
\rangle.
\end{align}
\end{widetext}
    
\end{section}

\begin{section}{Expression for \( S_{(ij)(kl)(mn)(pq),i_1j_1k_1} \) }
\label{app:S}

By using Eq.\eqref{Eq3.5}, the reduction formula Eq.\eqref{Eq6.4} for the fourth-order quantity \( \langle A_{(ij)}A_{(kl)}A_{(mn)}A_{(pq)}\rangle\) in the RPA basis can be directly evaluated, with
\begin{widetext}
\begin{align}
S_{(ij)(kl)(mn)(pq),i_1j_1k_1}
={}&\quad
 \Pm{ij}{n k_1}\,\Pm{kl}{k_1 j_1}\,\Qm{pq}{i_1 m}
-\Pm{ij}{n j_1}\,\Pm{kl}{i_1 k_1}\,\Qm{pq}{k_1 m}
\nonumber\\
&+
 \Pm{ij}{k_1 j_1}\,\Pm{kl}{i_1 m}\,\Qm{pq}{n k_1}
-\Pm{ij}{i_1 k_1}\,\Pm{kl}{n j_1}\,\Qm{pq}{k_1 m}
\nonumber\\
&+
 \Pm{ij}{i_1 m}\,\Pm{kl}{k_1 j_1}\,\Qm{pq}{n k_1}
-\Pm{ij}{k_1 m}\,\Pm{kl}{i_1 k_1}\,\Qm{pq}{n j_1}
\nonumber\\
&+
 \Pm{ij}{k_1 j_1}\,\Qm{mn}{l k_1}\,\Qm{pq}{i_1 k}
-\Pm{ij}{i_1 k_1}\,\Qm{mn}{l j_1}\,\Qm{pq}{k_1 k}
\nonumber\\
&+
 \Pm{ij}{i_1 k}\,\Qm{mn}{k_1 j_1}\,\Qm{pq}{l k_1}
-\Pm{ij}{l j_1}\,\Qm{mn}{i_1 k_1}\,\Qm{pq}{k_1 k}
\nonumber\\
&+
 \Pm{ij}{k_1 j_1}\,\Qm{mn}{i_1 k}\,\Qm{pq}{l k_1}
-\Pm{ij}{i_1 k_1}\,\Qm{mn}{k_1 k}\,\Qm{pq}{l j_1}
\nonumber\\
&+
 \Qm{kl}{k_1 j_1}\,\Qm{mn}{j k_1}\,\Qm{pq}{i_1 i}
-\Qm{kl}{i_1 k_1}\,\Qm{mn}{j j_1}\,\Qm{pq}{k_1 i}
\nonumber\\
&+
 \Qm{kl}{i_1 i}\,\Qm{mn}{k_1 j_1}\,\Qm{pq}{j k_1}
-\Qm{kl}{j j_1}\,\Qm{mn}{i_1 k_1}\,\Qm{pq}{k_1 i}
\nonumber\\
&+
 \Qm{kl}{k_1 j_1}\,\Qm{mn}{i_1 i}\,\Qm{pq}{j k_1}
-\Qm{kl}{i_1 k_1}\,\Qm{mn}{k_1 i}\,\Qm{pq}{j j_1} \, .
\end{align}
\end{widetext}

\end{section}

\bibliography{PTAref}












\end{document}